\documentclass[11pt,a4paper]{article}
\usepackage[a4paper,centering,scale=0.7]{geometry}
\usepackage{amsmath,amsthm,amsfonts}
\usepackage{setspace}
\usepackage[infoshow]{tabularx}
\usepackage{graphicx}
\usepackage{color}
\usepackage[nofiglist, notablist]{endfloat}

\newcommand{\E}{\mathbb{E}}

\renewcommand{\P}{\mathbb{P}}
\newcommand{\erre}{\mathbb{R}}

\newcommand{\eqd}{\stackrel{d}{=}}
\newcommand{\ip}[2]{\langle #1,#2 \rangle}
\newcommand{\limd}{\Rightarrow}
\newcommand{\sgn}{\mathop{\mathrm{sgn}}}

\newtheorem{prop}{Proposition}
\newtheorem{lemma}[prop]{Lemma}

\theoremstyle{remark}
\newtheorem{rmk}[prop]{Remark}

\title{Multivariate heavy-tailed models for Value-at-Risk
  estimation}

\author{Carlo Marinelli\thanks{Facolt\`a di Economia, Universit\`a di Bolzano, Piazza Universit\`a 1, I-39100 Bolzano, Italy.} 
 \and Stefano d'Addona\thanks{Department of International Studies, University of Rome 3, Via G. Chiabrera,
 199, I-00145 Rome, Italy, e-mail \texttt{daddona@uniroma3.it}, and Baruch College CUNY, One Bernard Baruch Way, 10010 New York, USA.}
 \and Svetlozar T. Rachev\thanks{Department of Applied Mathematics \& Statistics, Stony Brook University, Stony Brook, NY 11794-3600, USA.}}

\date{\small 15 May 2010 \\ Revised 17 April 2011 and 2 September 2011}

\begin{document}
\maketitle

\newpage
\setcounter{page}{1}
\begin{center}
\LARGE{Multivariate heavy-tailed models for Value-at-Risk estimation}
\end{center}
\bigskip

\begin{abstract}
  For purposes of Value-at-Risk estimation, we consider several
  multivariate families of heavy-tailed distributions, which can be
  seen as multidimensional versions of Paretian stable and Student's
  $t$ distributions allowing different marginals to have different
  indices of tail thickness. After a discussion of relevant estimation
  and simulation issues, we conduct a backtesting study on a set of
  portfolios containing derivative instruments, using historical US
  stock price data.
\end{abstract}

\bigskip
\bigskip


\bigskip
\bigskip


\newpage
\setstretch{1.5}

%\noindent \textit{Keywords: Multidimensional stable-like distribution, 
%Multidimensional $t$-like distribution, tail thickness,
%tail dependence, Backtesting, Delta-Gamma approximation.} 
%\setstretch{1.5}

\section{Introduction}
The purpose of this paper is to assess the performance of some classes
of multivariate laws with heavy tails in the estimation of
Value-at-Risk for nonlinear portfolios. The inadequacy of Gaussian
laws, in one or several dimensions, to model the distribution of risk
factors, especially in view of applications to risk modeling, is
well-documented in the empirical literature (see e.g.
\cite{berko,DuPa-overview} and references therein). Here we
concentrate on models for risk factors that are multivariate
extensions of the classical $\alpha$-stable and Student's $t$
distributions. In particular, we consider multivariate laws whose
marginals may have different indices of tail thickness, and/or whose
structures allow for tail dependence (i.e., roughly speaking, extreme
movements of several risk factors may happen together).

Let us briefly recall how VaR is usually estimated for nonlinear
portfolios (i.e. for portfolios containing derivative instruments),
and what kind of improvements have been proposed. In the simplest
setting, one uses a linear approximations of losses with normally
distributed risk factors: denoting by $L$ the loss over a certain time
period, one sets $L \approx \ip{\Delta}{X}$, where $X \sim N(m,Q)$ is
a $d$-dimensional vector of Gaussian risk factors, $\Delta$ is an
element of $\erre^d$, and $\ip{\cdot}{\cdot}$ stands for the usual
scalar product of two vectors. Then $\ip{\Delta}{X}$ follows a
one-dimensional Gaussian distribution with mean $\ip{\Delta}{m}$ and
variance $\ip{Q\Delta}{\Delta}$, so that (an approximation of) VaR can
be obtained immediately. However, it is clear that such a scheme
suffers from two major weaknesses: the linear approximation is
inaccurate, as the payoff function of derivatives is usually highly
non-linear, and the hypothesis that random factors are Gaussian is
often inappropriate, as briefly mentioned above (the literature on
this issue is very rich -- see e.g. \cite{blatta,infame,fama}, to
mention just a few classical references).
Among the many improvements that have been suggested in literature,
some focus on a better modeling of the nonlinear relation between $L$
and $X$ (e.g. by using quadratic approximations of the type $L \approx
\ip{\Delta}{X}+\ip{\Gamma X}{X}$), but still assuming $X$ Gaussian
(see e.g. \cite{DuPa-overview}), while others introduce alternative
distributions of portfolio losses, often just in the univariate
setting (see e.g. \cite{Frain:08,rachev-libro}). To the best of our
knowledge, however, there are only a small number of studies devoted
to models that take into account both non-linearities and
non-normality in a multivariate setting: Duffie and Pan \cite{DuPa}
and Glasserman et al. \cite{hdp} adopt the quadratic approximation and
non-Gaussian risk factors. In particular, risk factors include a jump
component in the first work, and are modeled by multivariate $t$
distributions (or a modification thereof) in the latter. However, both
works are devoted to different issues (analytic approximations and
efficient simulation techniques, respectively), therefore they do not
address the statistical issues related to the implementation of their
models, and do not test their empirical performance on real data.

Our contributions are the following: we introduce a stable-like model
for risk factors obtained by multivariate subordination of a Gaussian
law on $\erre^d$ (see {\S}\ref{sec:stablike}), such that each marginal
(i.e. each risk factor) can have a different index of tail
thickness. We construct estimators for the parameters of this
distribution and we study their asymptotic behavior. An analogous
program is carried out for a multivariate $t$-like law (see
{\S}\ref{sec:mtv}).
In \S\S\ref{sec:metas}--\ref{sec:warped} we consider models of risk
factors obtained by ``deforming'' the marginals of symmetric
sub-Gaussian $\alpha$-stable and multivariate $t$-distributed random
vectors, respectively. Equivalently, using the language of copulas, we
consider models of risk factors with symmetric $\alpha$-stable
resp. $t$-distributed marginals (possibly with different parameters)
on which a sub-Gaussian $\alpha$-stable resp. multivariate $t$ copula is
superimposed.
In {\S}\ref{sec:becchete} we provide an extensive back-testing study
of all parametric families of distributions using real data, on
portfolios containing both standard and exotic options, relying both
on full revaluation of the portfolio value and on its quadratic
approximation.

\medskip

We conclude this introduction with a few words about notation: given a
(possibly random) $d$-dimensional vector $X$, we shall denote its
$i$-th component, $1 \leq i \leq d$, by $X_i$. The inverse of an
invertible function $f$ will be denoted by $f^\leftarrow$. The
Gaussian measure with mean $m$ and covariance matrix $Q$ will be
denoted $N(m,Q)$. We shall write $X \sim \mathcal{L}$, with $X$ a
random variable and $\mathcal{L}$ a probability measure, to mean that
the law of $X$ is $\mathcal{L}$. The $\alpha$-stable measure on the real
line with index $\alpha$, skewness $\beta$, scale $\sigma$ and
location $\mu$ is denoted by $S_\alpha(\sigma,\beta,\mu)$, and we
always use the parametrization adopted in \cite{ST}.

\subsection*{Acknowledgments}
We are very grateful to two anonymous referees for a careful reading
of previous versions of this paper. Their corrections and useful
suggestions led to an improved version and a better presentation of
our results.

\section{Multivariate stable-like risk factors}
\label{sec:stablike}
\subsection{Description of the model}
Given a probability space $(\Omega,\mathcal{F},\P)$, let
$X:\Omega\to\erre^d$ be a random vector of risk factors such that
\begin{equation}
\label{eq:X}
X=A^{1/2}G,
\end{equation}
where $A=\mathrm{diag}(A_1,\ldots,A_d)$ is a diagonal random matrix
with independent entries,
\begin{equation}
\label{eq:A}
A_i \sim S_{\alpha_i/2}\Big(\big(\cos{\pi\alpha_i\over
4}\big)^{2/\alpha_i},1,0\Big)
\quad \;\; \forall i=1,\ldots,d,
\end{equation}
and $G$ is a $\erre^d$-valued Gaussian random vector, independent of
$A$, with mean zero and covariance matrix $Q$. In (\ref{eq:A}) we
assume $\alpha_i\in ]1,2[$ for all $i=1,\ldots,d$.

Note that (\ref{eq:X}) and (\ref{eq:A}) imply that, for each
$i=1,\ldots,d$, the $i$-th marginal of $X$ has distribution
$S_{\alpha_i}(\sigma_i,0,0)$, where $\E[G_i^2]=2\sigma_i^2$. In
particular, risk factors are allowed to have different indices of tail
thickness $\alpha_i$, and they are dependent through the Gaussian
component $G$.

\subsection{Estimation}
\label{sec:estimation} 
\begin{figure}
\begin{center}
\includegraphics[width=0.75\textwidth]{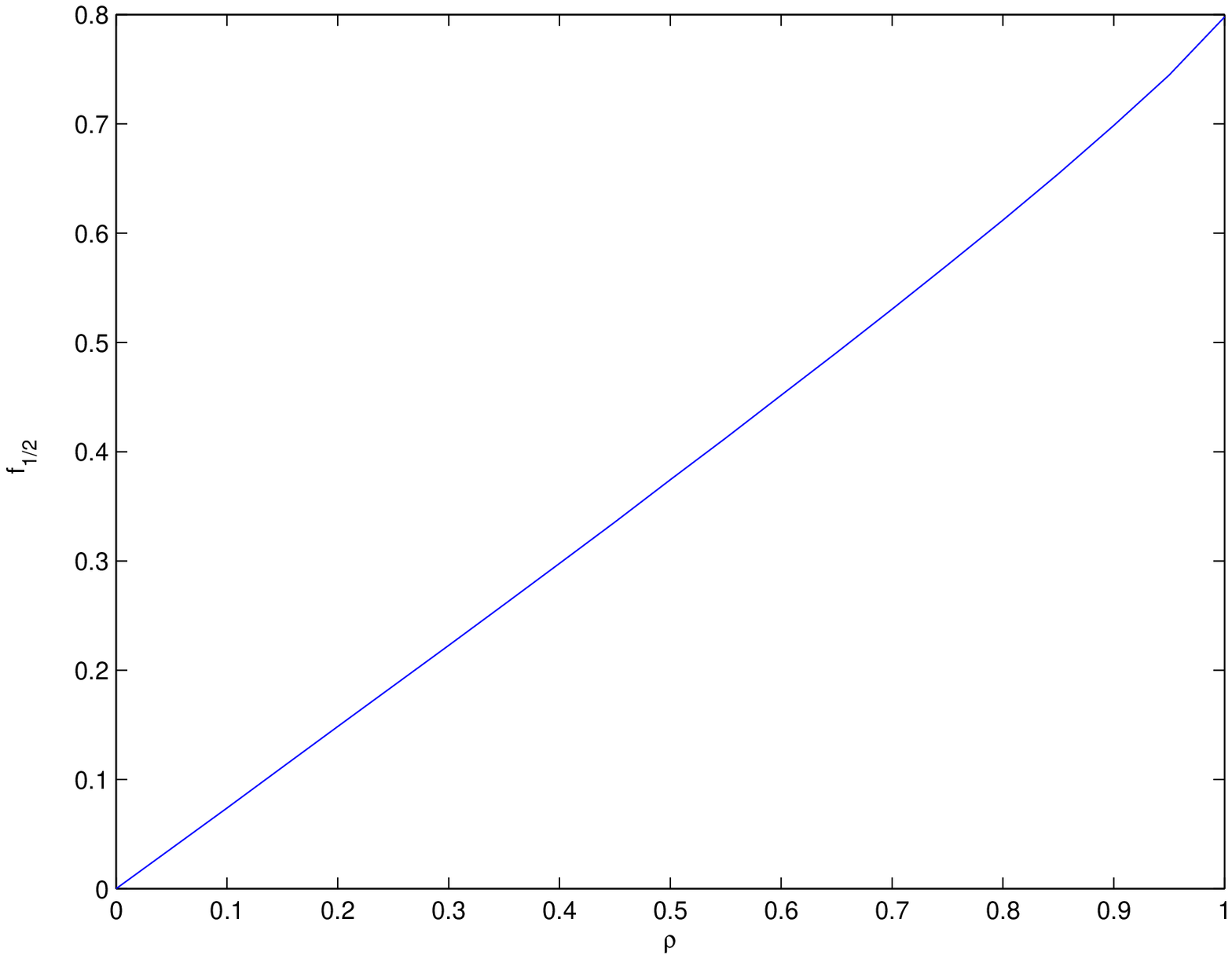}
\end{center}
\caption{Plot of the function $f_p$, with $p=1/2$.}
\label{fig:f}
\end{figure}

Let $X(t)$, $t=1,\ldots,n$ be independent samples from the
distribution of $X$. For $p < \min_{1\leq i\leq d}(\alpha_i)/2$,
define the (improper) sample $p$-th moment as
\[
M_p(n) = n^{-1}\sum_{t=1}^n 
X_i(t)^{\langle p \rangle}X_j(t)^{\langle p \rangle},
\]   
where $X^{\langle p\rangle}:=|X|^p\sgn(X)$.
Note that, by Cauchy-Schwarz' inequality, we have
\[
\E|X_iX_j|^p \leq \big(\E|X_i|^{2p}\big)^{1/2} \big(\E|X_j|^{2p}\big)^{1/2}
< \infty,
\]
thus also, by Kolmogorov's strong law of large numbers,
\[
\lim_{n\to\infty} M_p(n) = \E(X_iX_j)^{\langle p \rangle} \quad \textrm{a.s.}. 
\]
Since the random matrix $A$ and the random vector $G$ are independent, one has
\[
\E(X_iX_j)^{\langle p \rangle} = \Big(\E A_i^{p/2}\Big) \Big(\E A_j^{p/2}\Big)
\E(G_iG_j)^{\langle p \rangle},
\]
where (see e.g. \cite[p.~18]{ST})
\[ 
\E A_i^{p/2} = 
\frac{2^{p/2}\Gamma(1-p/\alpha_i)}{p \int_0^\infty u^{-p/2-1}\sin^2u\,du}\;
\Big( 1 + \tan^2 \frac{\alpha_i\pi}{4} \Big)^{p\over 2\alpha_i}
\Big(\cos{\pi\alpha_i\over 4}\Big)^{p\over\alpha_i}
\cos\frac{p\pi}{4}
=: C_{\alpha_i,p}.
\]
The constant $C_{\alpha,p}$ can be computed explicitly, recalling that
\[
\int_0^\infty u^{-p/2-1}\sin^2 u\,du =
-2^{p/2-1}\, \cos \frac{\pi p}{4}\, \Gamma(-p/2).
\]
Let us now define the function
\begin{align*}
  f_p:\; ]-1,1[ &\to \erre\\
           q &\mapsto \E[(Z_1Z_2)^{\langle p \rangle}],
\end{align*}
where $Z_1$, $Z_2$ are jointly normal random variables with covariance
matrix
\[
\left[\begin{array}{cc}
1 & q \\
q & 1 \\
\end{array}\right].
\]
For any given $p<\min_i(\alpha_i)/2$, matching the theoretical signed
$p$-th moments of $X_iX_j$ with their sample counterparts, we obtain
the following estimator for the matrix $Q$:
\[
\hat{Q}_{ij} = 2\sigma_i\sigma_j f_p^{\leftarrow} 
\left( { n^{-1}\sum_{t=1}^n X_i(t)^{\langle p \rangle}X_j(t)^{\langle p \rangle}
\over
2^p\sigma_i^p\sigma_j^p C_{\alpha_i,p} C_{\alpha_j,p} } \right),
\qquad i,\,j=1,\ldots,d.
\]
If $\{\sigma_i\}_i$ and $\{\alpha_i\}_i$ are not known a priori, but
we rather have only consistent estimators $\{\hat{\sigma}_{in}\}_i$ and
$\{\hat{\alpha}_{in}\}_i$, respectively, one can easily deduce (by
several applications of the continuous mapping theorem), that the
estimator of $Q$ obtained replacing $\alpha_i$ with
$\hat{\alpha}_{in}$ and $\sigma_i$ with $\hat{\sigma}_{in}$ in the
above expression is still consistent.
\begin{rmk}
  (i) As far as the estimation of the covariance matrix $Q$ is
  concerned, the heavy tailed assumption does not imply any extra
  computational burden.

  (ii) For our purposes, it is enough to choose $p=1/2$, as we always
  assume $\alpha_i>1$ for all $i=1,\ldots,d$ (as is well-known, this is
  equivalent to assuming that all returns have finite mean).

  (iii) Unfortunately we are not aware of an explicit expression for
  the function $q \mapsto f_p(q)$. However, it can be expressed
  as an integral with respect to a Gaussian measure in $\erre^2$:
  \begin{align}
    f_p(\rho) &= {1 \over 2\pi\sqrt{\det Q}}
    \int_{\erre^2} (x_1x_2)^{\langle p \rangle}
    e^{-\frac12\langle Q^{-1}x,x\rangle}\,dx \nonumber\\
    &= {1\over 2\pi\sqrt{1-q^2}}
    \int_{\erre^2} (x_1x_2)^{\langle p \rangle}
    e^{-{1\over2(1-q^2)}(x_1^2-2q x_1x_2+x_2^2)}\,dx_1\,dx_2
    \label{eq:integral}
  \end{align}
  which can be computed by numerical integration with essentially any
  accuracy. Figure \ref{fig:f} plots the function $f_{1/2}$ on the
  interval $[0,1[$.
\end{rmk}

Let us consider a simplified case: $d=2$,
$\sigma_1=\sigma_2=1/\sqrt{2}$, and $\alpha_1$, $\alpha_2$ given. The
assumption $d=2$ is harmless, as in any case the method works
componentwise. The case of unknown $\alpha_i$ and $\sigma_i$ can be
dealt with replacing them with their corresponding consistent
estimators, as discussed above.

Let us define
\begin{equation}
\label{eq:stim}
\hat{q}_n = f^{\leftarrow} 
\left( { n^{-1}\sum_{t=1}^n X_1(t)^{\langle p \rangle}X_2(t)^{\langle p \rangle}
\over C_{\alpha_1,p} C_{\alpha_2,p} } \right)
\end{equation}
We first prove the following lemma:
\begin{lemma}\label{lem:fp}
  The function $f_p:]-1,1[\to\erre$ is bounded, continuously
  differentiable, concave increasing on $]-1,0[$ and convex increasing
  on $]0,1[$.
\end{lemma}
\begin{proof}
  Boundedness follows by concavity of the function $x \mapsto |x|^p$ for
  $p<1$ and Jensen's inequality, that yield
  \[
  |f_p(q)| = |\E Z_1^{\langle p\rangle} Z_2^{\langle p\rangle}|
  \leq \E|Z_1Z_2|^p \leq (\E|Z_1Z_2|)^p \leq 1,
  \]
  where the last inequality follows by Cauchy-Schwarz' inequality and $\E
  Z_1^2=\E Z_2^2 =1$.
  Continuous differentiability w.r.t. $q$ is immediate by
  inspection of (\ref{eq:integral}).
  Differentiating (\ref{eq:integral}) w.r.t. $q$ twice, one gets
  (after some cumbersome but elementary calculations) $f'_p(q)>0$
  for all $q\in ]-1,1[$, and $f''_p(q)<0$ for $q<0$,
  $f''_p(0)=0$, $f''_p(q)>0$ for $q>0$. The lemma is thus
  proved.
\end{proof}

It is easy to prove that $\hat{q}_n$ is strongly consistent, i.e. that
$\hat{q}_n\to q$ a.s. as $n\to\infty$. In fact, as above, since
$p<(\min_i \alpha_i)/2$, by Kolmogorov's strong law of large numbers
one has
\[
f_p(\hat{q}_n) = { n^{-1}\sum_{t=1}^n X_1(t)^{\langle p \rangle}X_2(t)^{\langle p
\rangle} \over C_{\alpha_1,p} C_{\alpha_2,p} } 
\xrightarrow{n\to\infty} \E (Z_1Z_2)^{\langle p \rangle} = f_p(q)
\qquad \textrm{a.s.},
\]
from which we can conclude thanks to the continuous mapping theorem
and the continuity of $f^\leftarrow$.

\smallskip

We are now going to prove that the estimator (\ref{eq:stim}) is
asymptotically normal, under a more stringent assumption on the chosen
value of $p$. Let us define the function 
\begin{align*}
g_p: \erre^2 &\to \erre\\
           x &\mapsto
\frac{x_1^{\langle p \rangle}x_2^{\langle p \rangle}}{%
C_{\alpha_1,p}C_{\alpha_2,p}}.
\end{align*}
It is clear that the estimator (\ref{eq:stim}) can be defined as the
solution of the equation
\begin{equation}     \label{eq:furb}
\P_ng_p := \frac1n \sum_{k=1}^n g_p(X(k)) = \E_q g_p(X) =: f_p(q),
\end{equation}
where $\P_n$ stands for the (averaged) empirical measure of the sample
$X(1),\ldots,X(n)$, i.e.
\[
\P_n := \frac1n \sum_{k=1}^n \delta_{X(k)}.
\]
\begin{prop}
  If $p<\min_i(\alpha_i)/4$, then $\hat{q}_n$ is
  asymptotically normal and satisfies
  \begin{equation}     \label{eq:estimator}
  \sqrt{n}(\hat{q}_n-q) \limd
  N\Big( 0,f_p'(q)^{-2}\big(\E_q[g_p^2(X)]-f_p^2(q)\big) \Big),
  \end{equation}
  where ``$\limd$'' stands for convergence in law.
\end{prop}
\begin{proof}
  We have proved in lemma \ref{lem:fp} that $f_p(q)=\E_q g_p(X)$ is a
  bijection on the open set $]-1,1[$, it is continuously
  differentiable on its domain, and $f_p'(x)\neq 0$ for all $x\in
  ]-1,1[$.  Moreover, as it follows from (\ref{eq:stim}) and
  (\ref{eq:furb}), one can write
  \begin{equation}\label{eq:an}
  \sqrt{n}(\hat{q}_n-q) =
  \sqrt{n}\Big(f_p^{\leftarrow}(\P_ng_p)-f_p^{\leftarrow}(\E_qg_p(X))\Big).
  \end{equation}
  We have, by the strong law of large numbers, $\P_ng_p \to \E_qg_p$
  a.s. as $n\to\infty$. Recalling that by hypothesis
  $p<\min_i(\alpha_i)/4$, it follows that $\E_qg_p^2(X)<\infty$, hence,
  by the central limit theorem,
  \[
  \sqrt{n}(\P_ng_p - \E_qg_p(X)) \limd N(0,\E_qg_p^2(X)-f_p^2(q)).
  \]  
  An application of the delta method, taking into account the inverse
  function theorem, now yields the result.
\end{proof}
A shortcoming of the asymptotic confidence interval implied by the
above proposition is that the asymptotic variance depends on the
parameter to be estimated itself. One can overcome this problem by a
variance stabilizing transformation: let us define the function
$\gamma:]-1,1[ \to \erre$,
\[
\gamma_p(q) = \E_qg_p^2(X) - \big(\E_qg_p(X)\big)^2 =
\frac{\E|X_1X_2|^{2p}}{C_{\alpha_1,2p}C_{\alpha_2,2p}}
- \frac{\big( \E X_1^{\langle p \rangle} X_2^{\langle p \rangle} \big)^2}
{C^2_{\alpha_1,p}C^2_{\alpha_2,p}}
\]
and
\[
\varphi_p(x) = \int_0^x \frac{f_p'(y)}{\gamma_p^{1/2}(y)}\,dy.
\]
Then, again by the delta method, we obtain
\[
\sqrt{n}(\varphi_p(\hat{q}_n)-\varphi_p(q)) \limd
N\Big(0,\varphi'_p(q)^2\frac{\gamma_p(q)}{f_p'(q)^2}\Big) = N(0,1),
\]
and a corresponding asymptotic confidence interval for $q$ as
\[
q \in [\varphi_p^{\leftarrow}(\varphi_p(\hat{q}_n-z_\alpha/\sqrt{n}),
\varphi_p^{\leftarrow}(\varphi_p(\hat{q}_n+z_\alpha/\sqrt{n})].
\]
This asymptotic normality result for $\varphi_p(\hat{q}_n)$ would of
course be better if we had an explicity expression for $\varphi_p$,
which instead needs to be approximated numerically. However, since
both $f_p$ and $\gamma_p$ are smooth functions (i.e. at least $C^2$),
constructing a numerical approximation of $\varphi_p$ is a rather simple
task.
\begin{rmk}     \label{rmk:cave}
  The proof of the previous proposition, as well as the construction
  of the variance-stabilizing transformation, depend crucially on the
  assumption that $\sigma_1$, $\sigma_2$, $\alpha_1$, $\alpha_2$ are
  known (cf. the assumptions stated immediately before formula
  (\ref{eq:stim})). Therefore, for application purposes, the result
  should be applied with care, and would do only as a ``first
  approximation''. Nonetheless, it is also quite common in the
  estimation of multivariate models to separate the estimation of the
  parameters for the marginals from the estimation of the dependence
  structure.  It would certainly be interesting to obtain asymptotic
  confidence intervals treating simultaneously $\sigma_i$, $\alpha_i$,
  $i=1,2$, and $q$ as parameters to be estimated.
\end{rmk}

\subsection{Simulation}     \label{sec:simulation}
In view of the results of the previous subsection, we assume that the
covariance matrix $Q$ is known, hence, with a slight but harmless
abuse of notation, we shall write $Q$ instead of $\hat{Q}$.

Random vectors from the distribution of $X$ can be simulated
by the following simple algorithm:
\begin{itemize}
\item[(i)] generate $d$ independent random variables $Z_i \sim
  N(0,1)$, $i=1,\ldots,d$, and form the random vector
  $Z=(Z_1,\ldots,Z_d) \sim N(0,I)$, so that $Q^{1/2}Z \sim N(0,Q)$;
\item[(ii)] independently from $Z$, generate $d$ independent random
  variables from the distribution of $A_i$, $i=1,\ldots,d$, as defined
  in (\ref{eq:A});
\item[(iii)] setting $A=\mathrm{diag}(A_1,\ldots,A_d)$, one has that
  $A^{1/2}Q^{1/2}Z$ is a sample from the $d$-dimensional law of
  $X$
\end{itemize}

Note that the only computational overhead with respect to the
simulation of a Gaussian vector is the simulation of the stable
subordinators, for which nonetheless efficient algorithms are
available (see e.g. \cite{ST}).

\subsection{Extensions}
Let us remark that the model (\ref{eq:X}) for the vector of risk
factors can be extended to allow for asymmetries. In particular,
setting
\[
\tilde{X} = X + B = A^{1/2}G + B,
\]
where $B$ is a random vector, independent of $X$, with independent
components $B_i$ distributed according to the law
$S_{\alpha_i}(\sigma_{B_i},1,0)$, we have that the $i$-th marginal of
the vector $\tilde{X}$ has distribution
$S_{\alpha_i}(\tilde\sigma_i,\tilde{\beta}_i,0)$, where
\[
\tilde\sigma_i = (\sigma_i^{\alpha_i} + \sigma_{B_i}^{\alpha_i})^{1/\alpha_i},
\qquad
\tilde\beta_i = \frac{\sigma_{B_i}^{\alpha_i}}{\sigma_i^{\alpha_i}
                + \sigma_{B_i}^{\alpha_i}}.
\]
One can then estimate the parameters $\alpha_i$, $\tilde\sigma_i$ and
$\tilde\beta_i$ fitting (e.g. by maximum likelihood estimation) a
general Paretian stable law to observed data, and obtain
(corresponding estimates of the) values of $\sigma_i$, $\sigma_{B_i}$:
\[
\sigma_i = \big(1-\tilde{\beta}_i\big)^{1/\alpha_i} \tilde{\sigma}_i,
\qquad
\sigma_{B_i}= \tilde{\beta}_i^{1/\alpha_i} \tilde{\sigma}_i.
\]
Note also that, since we assume $\alpha_i>1$ for all $i=1,\ldots,d$,
we have $\E\tilde{X}_i = 0$ for all $i$. Finally, an estimate of $Q$
can be obtained by a rather involved modification of method of
fractional moments introduced in \S\ref{sec:estimation} above. In
particular, assuming $d=2$ for simplicity and using the notation of
\S\ref{sec:estimation}, let us set
\[
f_p(\rho; \alpha_1,\alpha_2,c_1,c_2) =
\E (c_1A_1G_1^{1/2}+B_1)^{\langle p \rangle}
   (c_1A_1G_1^{1/2}+B_2)^{\langle p \rangle},
\]
where $c_1$, $c_2$ are positive constants and $B_1$, $B_2$ are
independent with $B_i \sim S_{\alpha_i}(1,1,0)$, $i=1,2$. Then a
moment estimator for $\rho$ can be constructed in a rather obvious way
by matching theoretical and sample fractional moments using the
function $f_p(\cdot;\alpha_1,\alpha_2,c_1,c_2)$, treating $\alpha_i$,
$c_i$, $i=1,2$, as ``known'' (by an immediate scaling argument, the
constants $c_1$, $c_2$ are uniquely determined by $\sigma_i$ and
$\tilde{\sigma}_i$, $i=1,2$). This method performs unfortunately much
slower in comparison to the corresponding procedure in the symmetric
case, as the function $f_p$ cannot be computed off-line as in
\S\ref{sec:estimation}.

\smallskip

Moreover, model (\ref{eq:X}) does not allow for tail dependence among
different risk factors. As a remedy, one may use the series
representation of stable subordinators (see e.g. \cite{ST}), setting
\[
A_i = \sum_{k=0}^\infty \gamma_k^{2/\alpha_i},
\qquad i=1,\ldots,n,
\]
where $(\gamma_k)_{k\geq 0}$ is a (fixed) sequence of independent
standard Gamma random variables. The analysis of this model, however,
is considerably more involved, and we plan to elaborate on these
issues in a future work.

\section{Multivariate $t$-like risk factors}
\label{sec:mtv}
\subsection{Description of the model}
On a probability space $(\Omega,\mathcal{F},\mathbb{P})$, let us
consider a $d$-dimensional random vector of risk factors $X$ such that
\begin{equation}     \label{eq:mtv}
X_k = \frac{G_k}{\sqrt{V_k/\nu_k}},
\qquad k=1,\ldots,d,
\end{equation}
where $G \sim N(0,Q)$ and $V_1,\ldots,V_d$ are independent
one-dimensional $\chi^2$-distributed random variables with parameters
$\nu_1,\ldots,\nu_d$, respectively. We also assume that $G$ and
$(V_1,\ldots,V_d)$ are independent. Then, for each $k=1,\ldots,d$, the
$k$-th marginal of $X$ is distributed according to a Student's $t$ distribution
with parameter $\nu_k$, multiplied by $\sigma_k:=(\E G_k^2)^{1/2}$. In
particular, as in the case of the previous section, risk factors may
have different indices of tail thickness (measured by $\nu_k$), and
their dependence comes from the Gaussian component $G$.

\subsection{Estimation}
Assuming for the time being $\nu_k$, $k=1,\ldots,d$, to be known, let
us estimate the covariance matrix $Q$ by the method of moments. We
shall assume from now on that $\nu_k>2$ for all $k$, which implies in
particular that $\E X_k^2<\infty$ for all $k$. One has
\begin{align*}
\E X_h X_k &= \sqrt{\nu_h\nu_k} \, \E G_hG_k \, \E V_h^{-1/2} \, \E V_k^{-1/2}\\
&= Q_{hk} \, \sqrt{\nu_h\nu_k} \, \E V_h^{-1/2} \, \E V_k^{-1/2}
\end{align*}
for all $h \neq k$, and
\[
\E X_k^2 = Q_{kk} \, \nu_k \, \E V_k^{-1} = \sigma^2_k \, \nu_k \, \E V_k^{-1}.
\]
Denoting, for simplicity, a random variable with $\chi^2(\nu)$
distribution by $V$, the density of $V$ is given by
\[
f_\nu(x) = \frac{1}{2^{\nu/2}\Gamma(\nu/2)} x^{\nu/2-1} e^{-\frac{x}{2}},
\]
so that
\begin{align*}
\E V^{-1/2} &= \int_0^\infty x^{-1/2} f_\nu(x)\,dx
= \frac{1}{2^{\nu/2}\Gamma(\nu/2)} 
\int_0^\infty x^{\nu/2-3/2} e^{-\frac{x}{2}}\,dx\\
&= \frac{\Gamma(\nu/2-1/2)}{\sqrt{2}\,\Gamma(\nu/2)}
\end{align*}
and, similarly,
\[
\E V^{-1} = \int_0^\infty x^{\nu/2-2} e^{-\frac{x}{2}}\,dx
= \frac12 \frac{\Gamma(\nu/2-1)}{\Gamma(\nu/2)}
= \frac{1}{\nu-2}.
\]
Here we have used the definition of Gamma function,
\[
\Gamma(z) = \int_0^\infty t^{z-1} e^{-t}\,dt,
\qquad z>0,
\]
and its ``factorial'' property $\Gamma(z+1)=z\Gamma(z)$.
The above calculations yield
\[
Q_{hk} = \frac{2}{\sqrt{\nu_h\nu_k}} \,
\frac{\Gamma(\nu_h/2)}{\Gamma(\nu_h/2-1)} \,
\frac{\Gamma(\nu_k/2)}{\Gamma(\nu_k/2-1)} \,
\E X_hX_k,
\qquad h \neq k,
\]
and
\[
Q_{kk} = \sigma_k^2 = \frac{\nu_k-2}{\nu_k} \, \E X_k^2.
\]
We have thus obtained the following moment estimator for $Q$:
\[
\hat{Q}_{hk} = \frac{2}{\sqrt{\nu_h\nu_k}} \,
\frac{\Gamma(\nu_h/2)}{\Gamma(\nu_h/2-1)} \,
\frac{\Gamma(\nu_k/2)}{\Gamma(\nu_k/2-1)} \,
\frac{1}{n}\sum_{t=1}^n X_h(t)X_k(t),
\qquad h \neq k,
\]
and
\[
\hat{Q}_{kk} = \hat{\sigma}_k^2 = \frac{\nu_k-2}{\nu_k} \,
\frac{1}{n} \sum_{t=1}^n X_k(t)^2.
\]
Note that, for each $k$, $\nu_k$ can be estimate by one-dimensional
maximum likelihood on the $k$-th marginal, thus obtaining a family of
consistent estimators $\hat{\nu}_k$, $k=1,\ldots,d$. Therefore, the
corresponding estimator of $Q$ obtained by substituting in the
previous expressions each $\nu_k$ with $\hat{\nu}_k$, for each $k$,
is still consistent.

We can now prove that $\hat{Q}_{hk}$ is asymptotically normal. For
compactness of notation, we shall set
\[
C_\nu := \frac{\sqrt{2}}{\sqrt{\nu}}
\frac{\Gamma(\nu/2)}{\Gamma(\nu/2-1)},
\]
and we shall consider only the case $h \neq k$. The asymptotic
normality of the estimators $\hat{\sigma}_k$ can be established
analogously (see also {\S}\ref{ssec:mtv-ext}).
\begin{prop}
  Let $d=2$,
  \[
  Q = \left[\begin{array}{cc}
      1 & q \\
      q & 1 \\
  \end{array}\right],
  \]
  and
  \[
  \hat{q}_n := C_{\nu_1} C_{\nu_2} \, \frac{1}{n}\sum_{t=1}^n
  X_1(t)X_2(t).
  \]
  Then one has
  \[
  \sqrt{n}\big(\hat{q}_n - q\big) \limd N\big( 0, v_{\nu_1,\nu_2}(q) \big),
  \]
  where
  \[
  v_{\nu_1,\nu_2}(q) = \frac{\nu_1C_{\nu_1}^2}{\nu_1-2}
  \frac{\nu_2C_{\nu_2}^2}{\nu_2-2}(2q^2+1)-q^2
  \]
\end{prop}
\begin{proof}
  We have $\mathrm{Var}\,q_n = \E q_n^2 - q^2$ and
  \begin{align*}
  \E \hat{q}_n^2 &=  C_{\nu_1}^2 C_{\nu_2}^2 \, \E X_1^2 X_2^2 =
  \nu_1\nu_2 C_{\nu_1}^2 C_{\nu_2}^2 \, \E G_1^2 G_2^2
  \, \E V_1^{-1} \, \E V_2^{-1} \\
  &= \frac{\nu_1C_{\nu_1}^2}{\nu_1-2} \frac{\nu_2C_{\nu_2}^2}{\nu_2-2}
     \, \E G_1^2 G_2^2,
  \end{align*}
  where we have used the identity $\E V^{-1}=(\nu-2)^{-1}$. To compute
  $\E G_1^2 G_2^2$, let us write
\[
\begin{bmatrix}
  G_1 \\ G_2
\end{bmatrix}
\eqd
\begin{bmatrix}
  1 & 0 \\
  q & \sqrt{1-q^2}
\end{bmatrix}
\begin{bmatrix}
  Z_1 \\ Z_2
\end{bmatrix},
\]
with $(Z_1,Z_2) \sim N(0,I)$. This yields, recalling that the fourth
moment of a standard Gaussian measure is equal to 3,
\begin{equation}     \label{eq:gigia}
\E G_1^2 G_2^2 = q^2 \E Z_1^4 + 1-q^2 = 2q^2+1,
\end{equation}
thus also
\[
\mathrm{Var}\,\hat{q}_n = \frac{\nu_1C_{\nu_1}^2}{\nu_1-2}
\frac{\nu_2C_{\nu_2}^2}{\nu_2-2}
(2q^2+1)-q^2,
\]
whence the result follows by the central limit theorem.
\end{proof}
\begin{rmk}
  One could derive from this asymptotic normality result an asymptotic
  confidence interval using a variance stabilizing transformation, as
  shown in the previous section. The same caveats discussed in Remark
  \ref{rmk:cave} apply of course in this case as well.
\end{rmk}

\subsection{Simulation}
Generating random vectors from the distribution of a multivariate
$t$-like distribution is a straightforward modification of the
procedure outlined in {\S}\ref{sec:simulation} above.

\subsection{Extensions}     \label{ssec:mtv-ext}
Since marginals of the random vector $X$ follow a univariate $t$
distribution, they are symmetric. In order to allow for asymmetric
marginals, one may posit $X=(X_1,\ldots,X_d)$,
\[
X_k := \tilde{X}_k - \eta_k := \frac{G_k+m_k}{\sqrt{V_k/\nu_k}} - \eta_k,
\qquad k=1,\ldots,d,
\]
where $G \sim N(0,Q)$, and $m=(m_1,\ldots,m_d)$,
$\eta=(\eta_1,\ldots,\eta_d) \in \erre^d$. Then for the $k$-th
marginal one has that $X_k+\eta_k$ follows a noncentral
$t$-distribution. The reason for subtracting the vector $\eta$ from
$\tilde{X}$ is that $\E\tilde{X} \neq 0$, unless $m=0$, and it is
common to assume that risk factors have mean zero.  Unfortunately the
density of the noncentral $t$ law is expressed in terms of a definite
integral depending on parameters (see e.g. \cite{AubGaw03}), hence
maximum likelihood estimation on the marginals becomes numerically
quite involved. On the other hand, assuming $\nu_k>4$ for all $k$, one
can use the method of moments to construct estimators for
$\nu=(\nu_1,\ldots,\nu_d)$, $m$, $\eta$ and $Q$. In fact, considering
$k$ fixed and equal to 1 for the sake of simplicity, the constraint $\E
X_1=0$ translates into the relation
\[
\eta_1 = m_1 \sqrt{\nu_1} \, \E V_1^{-1/2} = m_1 \sqrt{\nu_1} \,
\frac{\Gamma((\nu_1-1)/2)}{\sqrt{2}\,\Gamma(\nu_1/2)}.
\]
Since we need to estimate four parameters, we need other three
equations. These can be obtained by matching the second, third, and
fourth sample moments to the corresponding theoretical moments, which
are known in closed form (see e.g. \cite{HPW61}).

We should also observe that in general it is not necessary to match
moments of integer order to obtain consistent and asymptotically
normal estimators. One may also use fractional moments, as it has been
done in the previous section, thus relaxing the assumptions on the
parameters $\nu_k$. For instance, let $X$ be as in (\ref{eq:mtv}),
$d=2$, $Q=\big[\begin{smallmatrix} 1&q \\ q&1\end{smallmatrix}\big]$,
and consider the problem of estimating $q$. Setting
$g_p(x_1,x_2)=x_1^{\langle p \rangle}x_2^{\langle p \rangle}$, we can write
\[
\E g_p(X) = (\nu_1\nu_2)^{p/2} \E V_1^{-p/2} \E V_2^{-p/2}
\E (G_1G_2)^{\langle p \rangle}.
\]
Note that $\E (G_1G_2)^{\langle p \rangle}=f_p(q)$, where $f_p$ is the
function introduced and studied in {\S}\ref{sec:estimation}, and, in
analogy to a calculation already encountered in this section,
\begin{equation}     \label{eq:putt}
\E V_k^{-p/2} = \frac{\Gamma(\nu_k/2-p/2)}{2^{p/2}\Gamma(\nu_k/2)},
\qquad k=1,2.
\end{equation}
This relation can be used as a basis for a moment estimator, as in
{\S}\ref{sec:estimation}. Choosing $p$ small enough, one does not need
to assume $\nu_k>2$.

\section{Multivariate meta-stable risk factors}
\label{sec:metas}

\subsection{Description of the model}
On a probability space $(\Omega,\mathcal{F},\P)$, let $G \sim N(0,Q)$
be a $d$-dimensional random vector with $\det Q \neq 0$, and
\[
A \sim S_{\alpha_0/2}\Big(
\big(\cos\frac{\pi\alpha_0}{4}\big)^{2/\alpha_0},1,0\Big),
\]
with $A$ and $G$ independent. The random vector $X':=A^{1/2}G$
is then symmetric $\alpha$-stable with characteristic function
\[
\E e^{i\ip{\xi}{X'}} = e^{-|\ip{Q\xi}{\xi}|^{\alpha_0/2}},
\]
in particular $X'$ has an elliptically contoured distribution (see
e.g. \cite{ST} for the properties of so-called sub-Gaussian
$\alpha$-stable laws, and \cite{Camb81} for elliptically contoured
distributions). As is well-known, the marginals of $X'$ are
$\alpha$-stable with index $\alpha_0$, i.e. they have all the same
index of tail thickness (as measured by $\alpha_0$). In order to build
a model allowing for different tail behavior along different
coordinates, one may set $X=f(X)$, with $f$ a deterministic
(nonlinear) injective function, for instance to ``deform'' the
marginals of $X'$ (a large part of the literature on the applications
of copulas to risk management is centered around this simple idea). A
common procedure (see e.g. \cite{hdp} in a slightly different context)
is to define a random vector $X=(X_1,\ldots,X_d)$ as
\begin{equation}     \label{eq:metas}
X_k = \sigma_k F_{\alpha_k}^\leftarrow(F_{\alpha_0}(X'_k)),
\qquad k=1,\ldots,d,
\end{equation}
with $\sigma_k$, $k=1,\ldots,d$ positive scaling constants, $\alpha_k
\in ]1,2]$ for all $k=1,\ldots,d$, and the diagonal elements of $Q$
are normalized to one. Here and throughout this section $F_\alpha$,
$\alpha \in ]0,2]$, stands for the one-dimensional distribution
function of a standard symmetric stable law with index $\alpha$. It is
clear that the law of the $k$-th marginal of $X$ is symmetric
$\alpha$-stable with index $\alpha_k$. Using the terminology introduced
in \cite{FFK02}, the random vector $X$ has a meta-elliptical
distribution, which we call meta-stable. Note that $X'$, hence also
$X$, are expected to have nontrivial tail dependence between any two
marginals because of the common factor $A$.

\subsection{Estimation}
One can estimate the parameters of a meta-stable distribution thanks to
the following remarkable relation (see \cite[Thm.~3.1]{FFK02} and also
\cite{Lind03}): let $X=(X_1,\ldots,X_d)$ be a random vector with
meta-elliptical distribution, and denote the Kendall's $\tau$
of $X_i$ and $X_j$, $i,j=1,\ldots,d$, by $\tau_{ij}$. Then we have
\[
\tau_{ij} = \frac{2}{\pi} \arcsin Q_{ij},
\]
which immediately yields the estimator
\[
\hat{Q}_{ij} = \sin \frac{\pi}{2} \hat{\tau}_{ij}.
\]
It is worth recalling that Kendall's tau statistic is a $U$-statistic
of order 2 with a bounded kernel, therefore it is asymptotically
normal (see e.g. \cite[\S12.1]{vdv}). Unfortunately however there does
not seem to be an explicit expression for the asymptotic variance, at
least not (to the best of our knowledge) in the cases considered in
this paper. One can also infer, by an application of the delta method,
that the above estimator of $Q_{ij}$, $i,\,j = 1,\ldots,d$, is also
asymptotically normal.

Moreover, the parameters $\sigma_k$ and $\alpha_k$, $k=1,\ldots,d$,
can easily be estimated e.g. by maximum likelihood on the marginals of
$X$. Finally, the parameter $\alpha_0$ can be estimated as follows,
where, in view of the above, we treat the parameters $\delta_k$,
$\nu_k$ and the matrix $Q$ as known (in practice they will have to be
replaced by their consistent estimators): defining the
$\erre^d$-valued random vector $U=(U_1,\ldots,U_d)$ as
\[
U_k := F_{\alpha_k}(X_k/\sigma_k), \qquad k=1,\ldots,d,
\]
since $U_k=F_{\alpha_0}(X'_k)$ for all $k=1,\ldots,d$, by the results
in Appendix \ref{sec:apetta}.1 we have that the law of $U$ admits the
density
\[
p_U(u_1,\ldots,u_d;\alpha_0) = \frac{%
h(F_{\alpha_0}^\leftarrow(u_1),\ldots,F_{\alpha_0}^\leftarrow(u_d))}{%
\prod_{k=1}^d f_{\alpha_0}(F_{\alpha_0}^\leftarrow(u_k))},
\]
where $f_{\alpha_0}$ is the density of $S_{\alpha_0}(1,0,0)$, and $h$
is the function defined in Appendix \ref{sec:apetta}.2. The parameter
$\alpha_0$ can then be estimated by maximum likelihood.

\subsection{Simulation}     \label{sec:metassim}
By (\ref{eq:metas}) we have $X_k=\sigma_k
F^\leftarrow_{\alpha_k}(F_{\alpha_0}(A^{1/2}G_k)$, $k=1,\ldots,d$,
from which a simulation scheme completely analogous to that outlined
in {\S}\ref{sec:simulation} can be devised. From the computational
point of view, the main problem is that there is no closed-form
expression for the cumulative distribution function and for the
inverse distribution function of a one-dimensional stable
law. However, there exist representations of them as integrals, which
can be implemented numerically. This procedure can become
computationally very expensive when simulating large samples. To
reduce the simulation time, one can compute off-line $F_\alpha(x)$ and
$F^\leftarrow_\alpha(x)$ for sufficiently many values of $\alpha$ and
$x$, and replace the numerical integration by interpolation. Since
both $(x,\alpha) \mapsto F_\alpha(x)$ and $(x,\alpha) \mapsto
F_\alpha^\leftarrow(x)$ are smooth functions (at least for
$\alpha>1$), interpolated values provide accurate approximations to
numerical integrals.

\subsection{Extensions}
The model can easily be extended to accomodate asymmetric marginals:
it is enough to set
\[
X_k = H_k^\leftarrow(F_{\alpha_0}(A^{1/2}G_k)), \qquad k=1,\ldots,d,
\]
where $H_k$ is the cumulative distribution function of the general
asymmetric centered $\alpha$-stable law
$S_{\alpha_k}(\sigma_k,\beta_k,0)$. The estimation of this extended
model is completely analogous to the symmetric case discussed above,
with the only difference that the parameters $\beta_k$,
$k=1,\ldots,d$, will also have to be estimated. This can be
accomplished again by maximum likelihood estimation on the marginals.

\section{Multivariate meta-$t$ risk factors}
\label{sec:warped}

\subsection{Description of the model}
On a probability space $(\Omega,\mathcal{F},\P)$, let $G$ and $V$ be a
$d$-dimensional random vector with law $N(0,Q)$ and an independent
one-dimensional random variable with $\chi^2$ distribution with
$\nu_0$ degrees of freedom, respectively. We shall call the law of the
random vector $X' = G/\sqrt{V/\nu_0}$ a multivariate $t$ distribution
(with parameters $\nu_0$ and $Q$). There are other possible
multivariate generalizations of Student's $t$ distribution (see
e.g. \cite{KoNa04}), but we shall concentrate exclusively on this
definition, which seems to be the most widely used in financial
applications.

In complete analogy to the meta-stable model discussed in the previous
section, the marginals of $X$ have the same tail thickness (as
measured by $\nu_0$), but one expects nontrivial tail dependence
between any two marginals because of the common factor $V$. In order
to allow for different tail behavior along different coordinates, one
can proceed as in the previous section. In particular (see
e.g. \cite{hdp}), one may define the $d$-dimensional random vector $X$ as
\begin{equation}     \label{eq:coppi}
X_k = \delta_k F_{\nu_k}^{\leftarrow}(F_{\nu_0}(X'_k)), \qquad k=1,\ldots,d,
\end{equation}
where $\delta_k>0$ for all $k=1,\ldots,d$, $F_\nu$ denotes (here and
throughout this section) the distribution function of a
one-dimensional $t$ law with $\nu$ degrees of freedom, and $Q_{kk}=1$
for all $k=1,\ldots,d$. Then the $k$-th marginal is $t$ distributed
with $\nu_k$ degrees of freedom, thus overcoming the problem of having
all marginals with the same tail thickness. The distribution of the
random vector $X$ belongs to the class of meta-elliptical
distributions introduced in \cite{FFK02}. In the latter reference the
law of $X$ is called meta-$t$, terminology which we have borrowed
here.

\subsection{Estimation}
The estimation algorithm for the meta-$t$ model is completely
analogous to the one for the meta-stable model. In fact, since $X$, as
just recalled, has a meta-elliptical distribution, the matrix $Q$ can
be estimated thanks to its relationship with Kendall's tau already
mentioned in Subsection \ref{sec:metas}.2 above (cf. also
\cite{dema,Lind03} about parameter estimation for the $t$ copula).  As
in the previous Section, Kendall's tau statistics as well as the
corresponding estimator of $Q_{ij}$, $i,j=1,\ldots,d$, are
asymptotically normal.

Similarly, the parameters $\delta_k$ and $\nu_k$, $k=1,\ldots,d$, can
easily be estimated by maximum likelihood on the marginals of
$X$. Finally, the parameter $\nu_0$ can be estimated by maximum
likelihood: treating, for the sake of simplicity, the parameters
$\delta_k$, $\nu_k$, $k=1,\ldots,d$, and the matrix $Q$ as known, let
us define the $d$-dimensional random vector $U$ as
\begin{equation}     \label{eq:symtcop}
U_k := F_{\nu_k}(X_k/\delta_k), \qquad k=1,\ldots,d.
\end{equation}
Recalling the explicit expression for the density of a multivariate
$t$ (see e.g. \cite{KoNa04}), (\ref{eq:coppi}) and the result in
Appendix \ref{sec:apetta}.1 imply that the law of $U$ admits the
density
\begin{align*}
p_U(u_1,\ldots,u_d;\nu_0) &= \frac{1}{(\det Q)^{1/2}}
\frac{\Gamma((\nu_0+d)/2)\Gamma(\nu_0/2)^{d-1}}{\Gamma((\nu_0+1)/2)^d}
\big( 1+\nu_0^{-1}\ip{Q^{-1}\tilde{u}}{\tilde{u}} \big)^{-\frac{\nu_0+d}{2}}\\
&\qquad \prod_{k=1}^d \big( 1+\nu_0^{-1} \tilde{u}_k^2
                     \big)^{\frac{\nu_0+1}{2}},
\end{align*}
where $\tilde{u_k}:=F_{\nu_0}^\leftarrow(u_k)$. In practice, of course
one needs to replace $\nu_k$, $\delta_k$ and $Q$ with the estimates
obtained e.g. by the above methods. 

\subsection{Simulation}
Random samples from the distribution of $X$ can be generated by
a rather straightforward modification of the procedure outlined in
Subsection \ref{sec:metassim}. In fact, the distribution function of the
univariate $t$ distribution, as well as its inverse, are implemented
in several software packages (such as Octave), even though they do not
admit a closed-form representation.\footnote{It might be better to say
  that they do, but in terms of hypergeometric functions.}

\subsection{Extensions}
As in the meta-stable case, one can generalize meta-$t$ laws to allow
for skewed marginals replacing $F_{\nu_k}$, $k=1,\ldots,d$, in
(\ref{eq:symtcop}) with the cumulative distribution functions of
noncentral $t$ laws, in analogy to the case discussed in
\S\ref{ssec:mtv-ext}.

\section{Estimation of Value-at-Risk by simulation}
\label{sec:simul}
We shall denote by $L$ the loss of a portfolio depending on the vector
of risk factors $X$. Recall that the Value-at-Risk (VaR) of a
portfolio at confidence level $\beta$ (usually $\beta=0.95$ or
$\beta=0.99$) is simply the $\beta$ quantile of the distribution of
portfolio losses. Since it is in general very difficult, if not
impossible, to obtain analytically tractable expressions for the
distribution function of the random variable $L$ (even if the density
function, or the characteristic function, of the vector $X$ is known
in closed form), one usually estimates quantiles of $L$ by generating
random samples from its distribution and computing the corresponding
empirical quantiles. We shall exclusively deal with the so-called
parametric (estimated) VaR, in the sense that we fit to observed data
the parameters of a given family of distributions for the vector $X$
of random factors, and we generate random samples from the law of $X$.
In order to obtain a sample from the law of $L$ we should know the
functional relation between $L$ and $X$. For a linear portfolio
(roughly, a portfolio without derivative instruments), one simply has
$L=\langle w,X\rangle$, where $w \in \erre^d$. In
the more interesting case of a portfolio containing derivatives, one
has $L=f(X)$, where $f: \erre^d \to \erre$ is a nonlinear
function. Unless the derivatives in the portfolio are very simple, the
function $f$ may not admit a closed-form representation, or could just
be obtained by nontrivial numerical procedures, that would have to be
carried out for each random sample of $X$. For this reason one usually
relies on approximations of the function $f$ of the form
\[
L \approx f(0) + \langle f'(0),X \rangle
+ \frac12 \langle f''(0)X,X \rangle,
\]
which is obviously motivated by the second-order Taylor expansion of
the function $\erre^d \ni x \mapsto f(x)$ around zero. The values of
$f'(0)$ and $f''(0)$ are in general determined by the so-called greeks
(in this case, Delta, Gamma and Theta) of the derivatives in the
portfolio. Note that in the above approximation the possible
dependence of $f$ on time can be taken into account by including time
in the set of risk factors.

The analytic computation, or just approximation, of quantiles of
quadratic forms in random vectors (other than Gaussian) is in general
a very difficult task. Simulation is hence a viable alternative, as
long as one can generate samples from the distribution of $X$.

We are going to perform a backtesting study on the four classes of
parametric models for the distribution of risk factors introduced in
Sections 2-5, to which we refer for the corresponding estimation and
simulation procedures. Value-at-Risk is just estimated by empirical
quantiles of random samples of $L$, either obtained by full
revaluation, or by the above quadratic approximation. In particular,
we do not focus on efficient simulation methods for quantile
estimation, but we are rather interested on the relative performance
of different distributional hypotheses for risk factors, when tested
on real data.

Let us also recall that all parametric families of multivariate laws
that we fit to data are symmetric. A detailed comparison of the
empirical performance of symmetric models and (some of) their
asymmetric counterparts is outside the scope of the present paper, and
it is left as an interesting question for future work.

% data.tex

\section{Empirical tests}
\label{sec:becchete}

\subsection{The data set}
\label{sec:data}

We consider two portfolios of underlyings with quite different
characteristics: portfolio $A$ is more diversified, while portfolio
$B$ is strongly correlated. In particular, portfolio $A$ is composed
of two US stocks from each of four different industries, while
portfolio $B$ is composed of eight US stocks from a single
industry\footnote{The selected stocks are Apple, Bank of America,
  Chevron, Citigroup, Conoco, Microsoft, Johnson and Johnson, and
  Pfitzer for portfolio $A$ and American Express, Banco Santander,
  Bank of America, Barclays, Citigroup, JP Morgan Chase, U.S. Bancorp,
  and Wells Fargo for portfolio $B$.}.  While portfolio $A$ is, in
some sense, more realistic (e.g. from the point of view of an investor
aiming at holding a reasonably diversified portfolio), portfolio $B$
is constructed as a ``stress test'' portfolio with high tail thickness
and (potentially) high tail dependence.

The raw price series are freely available on the Internet, and the
returns are calculated as daily log-differences\footnote{We restrict
  ourselves to consider daily data for two reasons: the first and most
  important is that the industry and regulatory standard is to compute
  VaR and related risk measures on a daily basis. On the other hand,
  studying lower frequencies (such as weekly or monthly) would
  considerably decrease the size of our samples, possibly invalidating
  the asymptotic properties of the proposed estimators.}. The data set
covers the time period from 2-Jan-1991 through 31-Dec-2008.

Let us provide a few descriptive statistics of the data set. Table
\ref{tbl:financial_stats} displays the sample kurtosis for each stock
return.  Note that all values are (much) larger than 3, thus providing
(rough) empirical evidence of tail-thickness of the underlying
distribution. The corresponding adjusted Jarque-Bera test statistic
(see e.g. \cite{Urzua}), is reported in the last column of Table
\ref{tbl:financial_stats} ($p$-values are in parentheses): for each
time series the hypothesis of an underlying Gaussian distribution is
rejected at $1\%$ level.

%
%%%%%%%%%%%%%%%%%%%%%%%%%%DESCRIPTIVE TABLE
\begin{table}[t]
\caption{Descriptive statistics of financial series}
\label{tbl:financial_stats} {\footnotesize \vspace{4mm} % 
  \setstretch{1} %\raggedleft
\scriptsize This table reports the sample kurtosis and the adjusted Jarque and Bera test for the
log-returns of the analyzed time series.
\\
\vspace{4mm}
\begin{tabular*}{\textwidth}{@{\extracolsep{\fill}}lcc}
\hline
&  Kurtosis  & Adj. J\&B     \\
American Express &    9.084   & 7066 \\
&&$(0.00)$\\
Apple &    57.786   & 573091 \\
&&$(0.00)$\\
Banco Santander &   10.040 & 9452   \\ 
&&$(0.00)$\\
Bank of America &   25.918 & 99815   \\ 
&&$(0.00)$\\
Barclays &   15.267 & 28907   \\ 
&&$(0.00)$\\
Chevron &    13.491  &20904  \\
&&$(0.00)$\\
Conoco &      8.373   &5529 \\
&&$(0.00)$\\
Citigroup &   38.374  & 237526  \\ 
&&$(0.00)$\\
Johnson \& Johnson &    9.735  & 8627  \\ 
&&$(0.00)$\\
JPM Chase  &    11.255 & 12946   \\ 
&&$(0.00)$\\
Microsoft &    8.186  & 5105  \\
&&$(0.00)$\\
%Oracle &  0.001  &    0.034   &   0.176   &    12.145 & 15895   \\
%&&&&&$(0.00)$\\
Pfitzer &    5.914  & 1632  \\ 
&&$(0.00)$\\
%Symantec &  0.000  &    0.038   &   -0.896   &    16.734 & 36398   \\
%&&&&&$(0.00)$\\
U.S. Bancorp &    22.514  & 72736  \\ 
&&$(0.00)$\\
Wells Fargo   &    20.869  & 60866  \\
&&$(0.00)$\\
%Xerox &  0.000  &    0.028   &   -0.570   &    23.875  &82921  \\
%&&&&&$(0.00)$\\
\hline
\end{tabular*}
}
\end{table}

\subsection{Test portfolios}
\label{sec:portfolio}
For each of the two portfolios, we construct three investment
strategies adding to the basic linear portfolios containing only the
eight underlyings (in equally value-weighted proportions) the
following positions in options:
\begin{description}
\item[NLL] long 10 calls and 5 puts on each asset (``NonLinear
  Long'');
\item[NLS] short 5 calls and 10 puts on each asset (``NonLinear
  Short'');
\item[NLDC] short 10 down-and-out calls with barrier equal to $95\%$
  of the asset price, and short 5 cash-or-nothing put with cash payoff
  equal to the strike price (``NonLinear Down and Cash'').
\end{description}
All options are European, at-the-money, and with time to expiration
equal to 6 months.
The nonlinear part of the six test portfolios is synthetic, in the
sense that option prices, unlike stock prices, are computed on the
basis of the information available on the corresponding underlying and
on (a proxy for) the risk-free rate, using Black-Scholes formula for
standard call and put options, and its variants for barrier and binary
options\footnote{We provide formulas for prices and sensitivities of
  these exotic options in Appendix \ref{sec:ape}.}. Even though this
procedure is incompatible with the non-Gaussian distributional
assumptions we are going to test, this is nonetheless common practice
(see e.g. \cite{hdp} for a more thorough discussion of this issue).

\subsection{Backtesting}
\label{sec:results}
Let us now turn to the analysis of the performance of the four
parametric distributions for risk factors introduced above, when
applied to the (predictive, i.e. out-of-sample) estimation of
Value-at-Risk. More precisely, we fit each of the multivariate laws to
a subset of the time series of stock returns (using a rolling window
consisting of 250 observations), and we estimate the 0.95 and 0.99
quantiles of the distribution of losses by simulation, i.e. selecting
the corresponding empirical quantiles from a simulated sample. In
particular, once a random sample from the distribution of $X$ is
obtained, we translate it into a random sample from the distribution
of portfolio losses either by a full revaluation of the portfolio
value for each sample, or by the usual delta-gamma quadratic
approximation (see {\S}\ref{sec:simul}).  Let $[t-\ell,t]$ denote the
time period over which the parametric families of distributions are
estimated, where $\ell$ stands for the (fixed) length of the rolling
window. Denoting by
$\mathrm{VaR}_t$ the empirical quantiles of the simulated distribution
of losses (with risk factors fitted over $[t-\ell,t]$), we form the
statistic
\[
\xi_{t+1} = \text{sgn}^{+} (L_{t+1}-\mathrm{VaR}_t),
\]
for all $t \in [\ell,T]$, where $T$ denotes the length of the time
series, $L_t$ stands for the observed loss of portfolio value over the
period $[t-1,t]$, and where $\sgn^+ x = 1$ if $x>0$, and equals zero
otherwise.  This procedure produces a different set of $(\xi_t)_{\ell
  \leq t \leq T}$, for each combination of test portfolio, model for
risk factors, quantile level (95\% and 99\%), and portfolio
revaluation method (full vs. quadratic approximation).

To assess the accuracy of the VaR estimates, we perform a simple
Proportion of Failure (PoF) test (cf. \cite{Kupiec:test}), in analogy
to the classical likelihood-ratio test.  In particular, setting
\begin{equation}
\label{test}
\zeta = -2\log\left(
\frac{(1-\beta)^x \beta^{(T-\ell-x)}}{p^x(1-p)^{(T-\ell-x)}}
\right),
\end{equation}
where $\beta \in \{0.95,\,0.99\}$,
\[
x:= \sum_{t=\ell+1}^T \xi_t,
\qquad
p:= \frac{x}{T-\ell},
\]
one expects $\zeta$ to be asymptotically $\chi^2$ distributed with one
degree of freedom. Therefore, the corresponding VaR model can be
considered reliable with a $95\%$ confidence level if $\zeta<\zeta_0
\approx 3.84$.

\subsubsection{Portfolio $A$}
The results of the backtesting procedure with full revaluation and
with quadratic approximation for portfolio A are collected in Tables \ref{tbl:var},
\ref{tbl:var-meta} and Tables \ref{tbl:quad}, \ref{tbl:quad-meta},
respectively, where values of $x$ are in the first column, $p$ in the
second column, and $\zeta$ in the third column.  Note that we
included, for comparison, VaR estimates obtained under the assumptions
that risk factors are jointly Gaussian.

As one may expect, the benchmark Gaussian approach fails at 99\%
confidence level for all three test portfolios. On the other hand, as
far as VaR estimates at 95\% confidence level are concerned, the
Gaussian approach is still satisfactory. The same performance is
displayed by the multivariate $t$-like approach. The stable-like
approach instead is rejected by the PoF test only once. We may
therefore say that our tests suggest that, between the two models
constructed by multiplying the marginals of a Gaussian vector by a set
of independent random variables, the stable-like approach might be
preferable. It may also be tempting to infer that models with trivial
tail dependence cannot adequately be used to estimate the probability
of large losses of financial portfolios. As we shall see below, this
conjecture does not seem to be supported by other empirical results.
Meta-$t$ and meta-stable both perform well, as the corresponding VaR
estimates cannot be rejected for any one of the test portfolios. Since
the estimated values of $\nu_0$ and $\alpha_0$ were very large for
long portions of the time series, we tested also the performance of
``degenerate'' meta-$t$ and meta-stable models, corresponding to the
limiting cases $\nu_0=\infty$ and $\alpha_0=2$, respectively. As it is
well-known, these models correspond to certain nonlinear deterministic
transformations of Gaussian laws (or, equivalently, to laws with
$t$-distributed and $\alpha$-stable distributed marginals,
respectively, and a Gaussian copula). Somewhat surprisingly, these
``degenerate'' meta-$t$ and meta-stable models give very accurate
results on our test portfolios. Since these models do have trivial
tail dependence, it is impossible to conclude, at least not with the
data at hand, that models with nontrivial tail dependence should be
preferable. In other words, our empirical result seem to imply that,
for portfolios whose underlyings are not (jointly) strongly dependent,
the sophistication of models allowing for both heavy tails and tail
dependence might not be indispensable.

Completely analogous observations could be made for the estimates of
VaR obtained by the delta-gamma quadratic approximation of portfolio
losses, for which we refer to Tables \ref{tbl:quad} and \ref{tbl:quad-meta}. As in the
case of full revaluation, the Gaussian approach performs remarkably
well at the 95\% confidence level. In this respect, it is probably
worth recalling that obtaining the quantiles of a quadratic form in
Gaussian vectors is particularly simple and can be done with very
little computational effort. In this sense, the classical quadratic
approximation with Gaussian risk factors could still be regarded as a
useful tool.

\subsubsection{Portfolio $B$}
The empirical results in the previous subsection, as already observed,
do not offer a decisive argument in favor of models featuring both
heavy tails and non-trivial tail dependence. For this reason it is
interesting to perform a back-testing analysis on the ``stress test''
portfolio $B$, whose underlyings are (presumably) heavy tailed and
strongly dependent.

The results with full revaluation and with quadratic approximation for
portfolio B are collected in Tables \ref{tbl:var2},
\ref{tbl:var-meta2} and Tables \ref{tbl:quad2}, \ref{tbl:quad-meta2},
respectively, where we included again, for comparison purposes, VaR
estimates obtained under the assumptions that risk factors are jointly
Gaussian.

One can see immediately (cf. Tables \ref{tbl:var2} and
\ref{tbl:quad2}) that both the $t$-like and the stable-like models, as
well as the standard Gaussian model, are unreliable at the 99\%
confidence level for all portfolios, and even at the 95\% confidence
level for the test portfolio NLDC containing exotic options. This
could be interpreted as empirical evidence that these classes of
models, all of which have trivial tail dependence, are not adequate to
estimate the probability of large losses, especially for highly
nonlinear portfolios. It does not seem possible, however, to assert
that the meta-$t$ and meta-stable models, both of which feature heavy
tails and non-trivial tail dependence, are superior in terms of their
empirical performance to their degenerate counterparts (i.e. the
meta-$t$ and meta-stable models with $\nu_0=\infty$ and $\alpha_0=2$,
respectively), which allow heavy tails but no tail dependence. In fact
(cf. Tables \ref{tbl:var-meta2} and \ref{tbl:quad-meta2}), the
performance of all meta-$t$ and meta-stable models is comparable for
the vanilla portfolios NLL and NLS, independently of having
non-trivial dependence or not (with the exception of the meta-stable
model which is rejecte in one case, see Table \ref{tbl:quad-meta2}).
The picture changes drastically for the exotic portfolio NLDC, for
which the meta-$t$, meta-stable and degenerate meta-$t$ models behave
poorly. On the other hand, surprisingly, the degenerate meta-stable
model display a good performance at the 99\% level. It appears to be
very difficult, if not impossible, to give an explanation for this
observation.

\begin{table}[t]
\caption{Value-at-Risk backtesting (full revaluation)}
\label{tbl:var}
{\footnotesize \vspace{2mm} %
\setstretch{1} %\raggedleft
%%\scriptsize
This table reports the results of a Value-at-Risk backtesting with
full revaluation of the first portfolio (portfolio A).
Panels A and B report the results for the long and short portfolios,
respectively, while Panel C reports the results for the down-and-out
and cash-or-nothing portfolio. Values marked with an asterisk
indicate that the corresponding model is not reliable.

\smallskip

\begin{tabular*}{\textwidth}{@{\extracolsep{\fill}}lccc}
%\\ \hline \multicolumn{4}{c}{Panel A: LP} \\
%\hline
% & Violations  &    Percentage       &    LR                   \\
%$t$ dist$_{95\%}(\nu_0=3)$ &   292  &    6.81\%   &26.69$^{\ast}$   \\ 
%$t$ dist$_{99\%}(\nu_0=3)$ &   69  &    1.61\%   &13.57$^{\ast}$  \\ 
%$t$ dist$_{95\%}(\nu_0=5)$ &   244  &    5.69\%   &4.17$^{\ast}$   \\ 
%$t$ dist$_{99\%}(\nu_0=5)$ &   64  &    1.49\%   &9.13$^{\ast}$    \\ 
%$t$ dist$_{95\%}(\nu_0=7)$ &   232  &    5.41\%   &1.48    \\ 
%$t$ dist$_{99\%}(\nu_0=7)$ &   62  &    1.45\%   &7.57$^{\ast}$    \\ 
%$t$ dist$_{95\%}$ multi  &   218  &    5.08\%   &0.06    \\ 
%$t$ dist$_{99\%}$ multi  &   80  &    1.87\%   &25.86$^{\ast}$    \\ 
%Stable$_{95\%}$ &   216  &    5.03\%   &    0.01    \\ 
%Stable$_{99\%}$ &   61  &    1.42\%   &    6.84$^{\ast}$    \\ 
%Gaussian$_{95\%}$ &   217  &    5.06\%   &    0.03    \\ 
%Gaussian$_{99\%}$ &   82  &    1.91\%   &    28.45$^{\ast}$   \\ 
\hline \multicolumn{4}{c}{Panel A: NLL} \\
\hline
 & Violations  &    Percentage       &    LR                   \\
%Warped $t$$_{95\%}(\nu_0=3)$ &   220  &    5.13\%   &    0.15     \\ 
%Warped $t$$_{99\%}(\nu_0=3)$ &   43  &    1.00\%   &    0.00    \\ 
%Warped $t$$_{95\%}(\nu_0=5)$ &   213  &    4.97\%   &   0.01    \\ 
%Warped $t$$_{99\%}(\nu_0=5)$ &   46  &    1.07\%   &    0.22    \\ 
%Warped $t$$_{95\%}(\nu_0=7)$ &   208  &    4.85\%   &    0.20    \\ 
%Warped $t$$_{99\%}(\nu_0=7)$ &   44  &    1.03\%   &  0.03    \\ 
$t$-like$_{95\%}$      &   204  &    4.76\%   &0.54    \\ 
$t$-like$_{99\%}$      &   64  &    1.49\%   &9.13$^{\ast}$    \\ 
Stable-like$_{95\%}$ &   205  &    4.78\%   &    0.44    \\ 
Stable-like$_{99\%}$ &   66  & 1.54\%   &    10.81$^{\ast}$     \\ 
Gaussian$_{95\%}$ &   191  &    4.45\%   &   2.79    \\ 
Gaussian$_{99\%}$ &   61  &    1.42\%   &    6.84$^{\ast}$     \\ 
\hline \multicolumn{4}{c}{Panel B: NLS} \\
\hline
 & Violations  &    Percentage       &    LR                   \\
%Warped $t$$_{95\%}(\nu_0=3)$ &   235  &    5.48\%   &   2.02   \\ 
%Warped $t$$_{99\%}(\nu_0=3)$ &   35  &    0.82\%   &    1.56   \\ 
%Warped $t$$_{95\%}(\nu_0=5)$ &   226  &    5.27\%   &    0.65   \\ 
%Warped $t$$_{99\%}(\nu_0=5)$ &   39 &    0.91\%   &    0.36   \\ 
%Warped $t$$_{95\%}(\nu_0=7)$ &   224  &    5.22\%   &    0.44   \\ 
%Warped $t$$_{99\%}(\nu_0=7)$ &   42 &    0.98\%   &    0.02   \\ 
$t$-like$_{95\%}$      &   225  &    5.25\%   &0.54    \\ 
$t$-like$_{99\%}$      &   64  &    1.49\%   &9.13$^{\ast}$    \\ 
Stable-like$_{95\%}$ &   223  &    5.20\%   &    0.36   \\ 
Stable-like$_{99\%}$ &   49  &    1.14\%   &    0.84  \\ 
Gaussian$_{95\%}$ &   207  &    4.83\%   &   0.27    \\ 
Gaussian$_{99\%}$ &   63  &    1.47\%   &    8.33$^{\ast}$   \\ 
\hline \multicolumn{4}{c}{Panel C: NLDC} \\
\hline
 & Violations  &    Percentage       &    LR                   \\
%Warped $t$$_{95\%}(\nu_0=3)$ &   235  &    5.48\%   &    2.02     \\ 
%Warped $t$$_{99\%}(\nu_0=3)$ &   42  &    0.98\%   &    0.02    \\ 
%Warped $t$$_{95\%}(\nu_0=5)$ &   231  &    5.39\%   &   1.32    \\ 
%Warped $t$$_{99\%}(\nu_0=5)$ &   44  &    1.03\%   &    1.03    \\ 
%Warped $t$$_{95\%}(\nu_0=7)$ &   224  &    5.22\%   &    0.45    \\ 
%Warped $t$$_{99\%}(\nu_0=7)$ &   44  &    1.03\%   &  0.03    \\ 
$t$-like$_{95\%}$      &   201  &    4.68\%   &0.90    \\ 
$t$-like$_{99\%}$      &   59  &    1.35\%   &5.48$^{\ast}$    \\
Stable-like$_{95\%}$ &   207  &    4.83\%   &    0.27    \\ 
Stable-like$_{99\%}$ &   47  & 1.10\%   &    0.39     \\ 
Gaussian$_{95\%}$ &   212  &    4.94\%   &   0.28    \\ 
Gaussian$_{99\%}$ &   65  &    1.52\%   &    9.95$^{\ast}$     \\ 
\end{tabular*}
}
\end{table}
\begin{table}[t]
\caption{Value-at-Risk backtesting (full revaluation)}
\label{tbl:var-meta}
{\footnotesize \vspace{2mm} %
\setstretch{1} %\raggedleft
%%\scriptsize
This table is a continuation of Table \ref{tbl:var}. The same notation is
used here.

\smallskip

\begin{tabular*}{\textwidth}{@{\extracolsep{\fill}}lccc}
\hline \multicolumn{4}{c}{Panel A: NLL} \\
\hline
& Violations  &    Percentage       &    LR                   \\
Meta-$t_{95\%}$ &   200  &    4.66\%   &1.04   \\ 
Meta-$t_{99\%}$ &   46  &    1.07\%   &0.22  \\
Meta-stable$_{95\%}$  &   223  &    5.20\%   &0.36   \\ 
Meta-stable$_{99\%}$  &   48  &    1.12\%   &0.59   \\ 
Meta-$t_{95\%}$ $(\nu_0=\infty)$ &   200  &    4.66\%   &1.04   \\ 
Meta-$t_{99\%}$ $(\nu_0=\infty)$ &   52  &    1.21\%   &1.83    \\ 
Meta-stable$_{95\%}$ $(\alpha_0=2)$ &   197  &    4.59\%   &1.53    \\ 
Meta-stable$_{99\%}$ $(\alpha_0=2)$ &   51  &    1.19\%   &1.46    \\ 
\hline \multicolumn{4}{c}{Panel B: NLS} \\
\hline
 & Violations  &    Percentage       &    LR                   \\
Meta-$t_{95\%}$ &   217  &    5.06\%   &0.03   \\ 
Meta-$t_{99\%}$ &   48  &    1.12\%   &0.59  \\ 
Meta-stable$_{95\%}$  &   233  &    5.43\%   &1.65    \\ 
Meta-stable$_{99\%}$  &   40  &    0.93\%   &0.20    \\ 
Meta-$t_{95\%}$ $(\nu_0=\infty)$ &   210  &    4.90\%   &0.10   \\ 
Meta-$t_{99\%}$ $(\nu_0=\infty)$ &   50  &    1.17\%   &1.13    \\ 
Meta-stable$_{95\%}$ $(\alpha_0=2)$ &   217  &    5.06\%   &0.03   \\ 
Meta-stable$_{99\%}$ $(\alpha_0=2)$ &   39  &    0.91\%   &0.37    \\  
\hline \multicolumn{4}{c}{Panel C: NLDC} \\
\hline
 & Violations  &    Percentage       &    LR                   \\
Meta-$t_{95\%}$ &   219  &    5.11\%   &0.10  \\ 
Meta-$t_{99\%}$ &   48  &    1.12\%   &0.59  \\ 
Meta-stable$_{95\%}$  &   224  &    5.22\%   &0.45  \\ 
Meta-stable$_{99\%}$  &   39  &    0.91\%   &0.37   \\ 
Meta-$t_{95\%}$ $(\nu_0=\infty)$ &   220  &    5.13\%   &0.15   \\ 
Meta-$t_{99\%}$ $(\nu_0=\infty)$ &   53  &    1.24\%   &2.24    \\ 
Meta-stable$_{95\%}$ $(\alpha_0=2)$ &   208  &    4.85\%   &0.20   \\ 
Meta-stable$_{99\%}$ $(\alpha_0=2)$ &   36  &    0.84\%   &1.18   \\ 
\end{tabular*}
}
\end{table}
\begin{table}[t]
\caption{Value-at-Risk backtesting (quadratic approximation)}
\label{tbl:quad}
{\footnotesize \vspace{2mm} %
\setstretch{1} %\raggedleft
%%
%\scriptsize
This table reports the results of a Value-at-Risk backtesting with
quadratic approximation of the first portfolio (portfolio A).
Panels A and B report the results for the long and short portfolios,
respectively, while Panel C reports the results for the down-and-out
and cash-or-nothing portfolio. Values marked with an asterisk
indicate that the corresponding model is not reliable.

\smallskip

\begin{tabular*}{\textwidth}{@{\extracolsep{\fill}}lccc}
\\ 
\hline \multicolumn{4}{c}{Panel A: NLL} \\
\hline
 & Violations  &    Percentage       &    LR                   \\
%Warped $t$$_{95\%}(\nu_0=3)$ &   219  &    5.11\%   &    0.10     \\ 
%Warped $t$$_{99\%}(\nu_0=3)$ &   42  &    0.98\%   &    0.02    \\ 
%Warped $t$$_{95\%}(\nu_0=5)$ &   210  &    4.90\%   &   0.10    \\ 
%Warped $t$$_{99\%}(\nu_0=5)$ &   42  &    0.98\%   &    0.02    \\ 
%Warped $t$$_{95\%}(\nu_0=7)$ &   202  &    4.71\%   &    0.77    \\ 
%Warped $t$$_{99\%}(\nu_0=7)$ &   42  &    0.98\%   &  0.02    \\ 
$t$-like$_{95\%}$      &   203  &    4.73\%   &0.65    \\ 
$t$-like$_{99\%}$      &   63  &    1.47\%   &8.33$^{\ast}$    \\ 
Stable-like$_{95\%}$ &   204  &    4.75\%   &    0.54    \\ 
Stable-like$_{99\%}$ &   65  & 1.52\%   &    9.95$^{\ast}$     \\ 
Gaussian$_{95\%}$ &   188  &    4.38\%   &   3.56    \\ 
Gaussian$_{99\%}$ &   61  &    1.42\%   &    6.84$^{\ast}$     \\
\hline \multicolumn{4}{c}{Panel B: NLS} \\
\hline
 & Violations  &    Percentage       &    LR                   \\
%Warped $t$$_{95\%}(\nu_0=3)$ &   239  &    5.57\%   &   2.87   \\ 
%Warped $t$$_{99\%}(\nu_0=3)$ &   39  &    0.91\%   &    0.37    \\ 
%Warped $t$$_{95\%}(\nu_0=5)$ &   232  &    5.41\%   &    1.48    \\ 
%Warped $t$$_{99\%}(\nu_0=5)$ &   42 &    0.98\%   &    0.02   \\ 
%Warped $t$$_{95\%}(\nu_0=7)$ &   227  &    5.30\%   &    0.77   \\ 
%Warped $t$$_{99\%}(\nu_0=7)$ &   47 &    1.10\%   &    0.39   \\ 
$t$-like$_{95\%}$      &   228  &    5.32\%   &0.89    \\ 
$t$-like$_{99\%}$      &   66  &    1.54\%   &10.81$^{\ast}$    \\ 
Stable-like$_{95\%}$ &   232  &    5.41\%   &    1.48   \\ 
Stable-like$_{99\%}$ &   49  &    1.14\%   &    0.84  \\ 
Gaussian$_{95\%}$ &   215  &    5.01\%   &   0.00    \\ 
Gaussian$_{99\%}$ &   63  &    1.47\%   &    8.33$^{\ast}$   \\ 
\hline \multicolumn{4}{c}{Panel C: NLDC} \\
\hline
 & Violations  &    Percentage       &    LR                   \\
%Warped $t$$_{95\%}(\nu_0=3)$ &   231  &    5.39\%   &    1.32     \\ 
%Warped $t$$_{99\%}(\nu_0=3)$ &   41  &    0.96\%   &    0.08    \\ 
%Warped $t$$_{95\%}(\nu_0=5)$ &   223  &    5.20\%   &   0.36    \\ 
%Warped $t$$_{99\%}(\nu_0=5)$ &   43  &    1.00\%   &    0.00    \\ 
%Warped $t$$_{95\%}(\nu_0=7)$ &   215  &    5.01\%   &    0.00    \\ 
%Warped $t$$_{99\%}(\nu_0=7)$ &   44  &    1.03\%   &  0.03    \\ 
$t$-like$_{95\%}$      &   185  &    4.31\%   &4.44$^{\ast}$    \\ 
$t$-like$_{99\%}$      &   58  &    1.35\%   &4.85$^{\ast}$    \\ 
Stable-like$_{95\%}$ &   204  &    4.76\%   &    0.54    \\ 
Stable-like$_{99\%}$ &   45  & 1.05\%   &    0.10     \\ 
Gaussian$_{95\%}$ &   208  &    4.85\%   &   0.20    \\ 
Gaussian$_{99\%}$ &   64  &    1.49\%   &    9.13$^{\ast}$     \\ 
\end{tabular*}
}
\end{table}

%%%%%%%meta tables

%
%
\begin{table}[t]
\caption{Value-at-Risk backtesting (quadratic approximation)}
\label{tbl:quad-meta}
{\footnotesize \vspace{2mm} %
\setstretch{1} %\raggedleft
%%
%\scriptsize

This table is a continuation of Table \ref{tbl:quad}. The same notation is
used here.

\smallskip

\begin{tabular*}{\textwidth}{@{\extracolsep{\fill}}lccc}
\hline \multicolumn{4}{c}{Panel A: NLL} \\
\hline
& Violations  &    Percentage       &    LR                   \\
Meta-$t_{95\%}$ &   200  &    4.66\%   &1.04  \\ 
Meta-$t_{99\%}$ &   44  &    1.03\%   &0.03  \\ 
Meta-stable$_{95\%}$  &   219  &    5.11\%   &0.10   \\ 
Meta-stable$_{99\%}$  &   45  &    1.05\%   &0.10    \\
Meta-$t_{95\%}$ $(\nu_0=\infty)$ &   200  &    4.66\%   &1.04   \\ 
Meta-$t_{99\%}$ $(\nu_0=\infty)$ &   52  &    1.21\%   &1.83    \\ 
Meta-stable$_{95\%}$ $(\alpha_0=2)$ &   193  &    4.50\%   &2.32    \\ 
Meta-stable$_{99\%}$ $(\alpha_0=2)$ &   50  &    1.17\%   &1.13    \\  
\hline \multicolumn{4}{c}{Panel B: NLS} \\
\hline
 & Violations  &    Percentage       &    LR                   \\
Meta-$t_{95\%}$ &   225  &    5.25\%   &0.54  \\ 
Meta-$t_{99\%}$ &   50  &    1.17\%   &1.13  \\ 
Meta-stable$_{95\%}$  &   236  &    5.50\%   &2.22    \\ 
Meta-stable$_{99\%}$  &   42  &    0.98\%   &0.02   \\  
Meta-$t_{95\%}$ $(\nu_0=\infty)$ &   220  &    5.13\%   &0.15  \\ 
Meta-$t_{99\%}$ $(\nu_0=\infty)$ &   52  &    1.21\%   &1.83    \\ 
Meta-stable$_{95\%}$ $(\alpha_0=2)$ &   227  &    5.29\%   &0.77    \\ 
Meta-stable$_{99\%}$ $(\alpha_0=2)$ &   44  &    1.03\%   &0.03    \\ 
\hline \multicolumn{4}{c}{Panel C: NLDC} \\
\hline
 & Violations  &    Percentage       &    LR                   \\
Meta-$t_{95\%}$ &   216 &    5.04\%   &0.01  \\ 
Meta-$t_{99\%}$ &   46  &    1.07\%   &0.22  \\ 
Meta-stable$_{95\%}$  &   223  &    5.20\%   &0.36   \\ 
Meta-stable$_{99\%}$  &   41  &    0.96\%   &0.08  \\ 
Meta-$t_{95\%}$ $(\nu_0=\infty)$ &   212  &    4.94\%   &0.03   \\ 
Meta-$t_{99\%}$ $(\nu_0=\infty)$ &   49  &    1.14\%   &0.84    \\ 
Meta-stable$_{95\%}$ $(\alpha_0=2)$ &   200  &    4.66\%   &1.04    \\ 
Meta-stable$_{99\%}$ $(\alpha_0=2)$ &   35  &    0.82\%   &1.56   \\ 
\end{tabular*}
}
\end{table}

\begin{table}[t]
\caption{Value-at-Risk backtesting (full revaluation)}
\label{tbl:var2}
{\footnotesize \vspace{2mm} %
\setstretch{1} %\raggedleft
%%\scriptsize
This table reports the results of a Value-at-Risk backtesting with
full revaluation of the second portfolio (portfolio B).
Panels A and B report the results for the long and short portfolios,
respectively, while Panel C reports the results for the down-and-out
and cash-or-nothing portfolio. Values marked with an aster
\begin{tabular*}{\textwidth}{@{\extracolsep{\fill}}lccc}
\hline \multicolumn{4}{c}{Panel A: NLL} \\
\hline
 & Violations  &    Percentage       &    LR                   \\
$t$-like$_{95\%}$      & 217  &    5.06\%   &0.03  \\ 
$t$-like$_{99\%}$      &   72  &    1.68\%   &16.59$^{\ast}$   \\ 
Stable-like$_{95\%}$ &   232  &    5.41\%   &    1.48    \\ 
Stable-like$_{99\%}$ &   88  & 2.05\%   &    36.77$^{\ast}$     \\ 
Gaussian$_{95\%}$ &   208  &    4.85\%   &   0.20    \\ 
Gaussian$_{99\%}$ &   70  &    1.63\%   &    14.55$^{\ast}$     \\ 
\hline \multicolumn{4}{c}{Panel B: NLS} \\
\hline
 & Violations  &    Percentage       &    LR                   \\
$t$-like$_{95\%}$     &   217  &    5.06\%   &0.03     \\ 
$t$-like$_{99\%}$      &   59  &    1.37\%   &5.48$^{\ast}$   \\ 
Stable-like$_{95\%}$ &   230  &    5.36\%   &    1.17   \\ 
Stable-like$_{99\%}$ &   507 &    1.33\%   &    4.26$^{\ast}$  \\ 
Gaussian$_{95\%}$ &   203  &    4.73\%   &   0.65    \\ 
Gaussian$_{99\%}$ &   64  &    1.63\%   &    9.13$^{\ast}$   \\ 
\hline \multicolumn{4}{c}{Panel C: NLDC} \\
\hline
 & Violations  &    Percentage       &    LR                   \\
$t$-like$_{95\%}$      &   253  &    5.90\%   &6.93$^{\ast}$   \\ 
$t$-like$_{99\%}$      &   78  &    1.82\%   &23.37$^{\ast}$    \\
Stable-like$_{95\%}$ &   250  &    5.83\%   &    5.92$^{\ast}$    \\ 
Stable-like$_{99\%}$ &   68  & 1.59\%   &    12.62$^{\ast}$     \\ 
Gaussian$_{95\%}$ &   250  &    5.83\%   &   5.92$^{\ast}$ \\ 
Gaussian$_{99\%}$ &  85  &    1.98\%   &    32.50$^{\ast}$     \\ 
\end{tabular*}
}
\end{table}
\begin{table}[t]
\caption{Value-at-Risk backtesting (full revaluation)}
\label{tbl:var-meta2}
{\footnotesize \vspace{2mm} %
\setstretch{1} %\raggedleft
%%\scriptsize
This table is a continuation of Table \ref{tbl:var2}. The same notation is
used here.

\smallskip

\begin{tabular*}{\textwidth}{@{\extracolsep{\fill}}lccc}
\hline \multicolumn{4}{c}{Panel A: NLL} \\
\hline
& Violations  &    Percentage       &    LR                   \\
Meta-$t_{95\%}$ &   227  &    5.50\%   &2.22   \\ 
Meta-$t_{99\%}$ &   51  &    1.05\%   &0.10  \\
Meta-stable$_{95\%}$  &   236  &    5.17\%   &0.28   \\ 
Meta-stable$_{99\%}$  &   45  &    0.96\%   &0.08   \\ 
Meta-$t_{95\%}$ $(\nu_0=\infty)$ &   223  &    5.20\%   &0.36   \\ 
Meta-$t_{99\%}$ $(\nu_0=\infty)$ &   54  &    1.26\%   &2.69     \\ 
Meta-stable$_{95\%}$ $(\alpha_0=2)$ &   221  &    5.15\%   &0.21    \\ 
Meta-stable$_{99\%}$ $(\alpha_0=2)$ &   46  &    1.07\%   &0.22    \\ 
\hline \multicolumn{4}{c}{Panel B: NLS} \\
\hline
 & Violations  &    Percentage       &    LR                   \\
Meta-$t_{95\%}$ &   223  &    5.20\%   &0.36   \\ 
Meta-$t_{99\%}$ &   44  &    1.03\%   &0.03  \\ 
Meta-stable$_{95\%}$  &   241  &    5.62\%   &3.34    \\ 
Meta-stable$_{99\%}$  &   41  &    0.99\%   &0.08    \\ 
Meta-$t_{95\%}$ $(\nu_0=\infty)$ &   219  &    5.10\%   &0.10    \\ 
Meta-$t_{99\%}$ $(\nu_0=\infty)$ &   49  &    1.14\%   &0.84    \\ 
Meta-stable$_{95\%}$ $(\alpha_0=2)$ &   223  &    5.20\%   &0.36   \\ 
Meta-stable$_{99\%}$ $(\alpha_0=2)$ &   44  &    1.03\%   &0.03   \\  
\hline \multicolumn{4}{c}{Panel C: NLDC} \\
\hline
 & Violations  &    Percentage       &    LR                   \\
Meta-$t_{95\%}$ &   259  &    6.04\%   &9.18$^{\ast}$  \\ 
Meta-$t_{99\%}$ &   68  &    1.59\%   &12.62$^{\ast}$  \\ 
Meta-stable$_{95\%}$  &   266  &    6.20\%   &12.18$^{\ast}$  \\ 
Meta-stable$_{99\%}$  &   59  &    1.38\%   &5.48$^{\ast}$   \\ 
Meta-$t_{95\%}$ $(\nu_0=\infty)$ &   255  &    5.95\%   &7.65$^{\ast}$  \\ 
Meta-$t_{99\%}$ $(\nu_0=\infty)$ &   69  &    1.61\%   &13.57$^{\ast}$    \\ 
Meta-stable$_{95\%}$ $(\alpha_0=2)$ &   252  &    5.88\%   &6.59$^{\ast}$   \\ 
Meta-stable$_{99\%}$ $(\alpha_0=2)$ &   49  &    1.14\%   &0.84   \\ 
\end{tabular*}
}
\end{table}

\begin{table}[t]
\caption{Value-at-Risk backtesting (quadratic approximation)}
\label{tbl:quad2}
{\footnotesize \vspace{2mm} %
\setstretch{1} %\raggedleft
%%
%\scriptsize
This table reports the results of a Value-at-Risk backtesting with
quadratic approximation of the second portfolio (portfolio B).
Panels A and B report the results for the long and short portfolios,
respectively, while Panel C reports the results for the down-and-out
and cash-or-nothing portfolio. Values marked with an asterisque
indicate that the corresponding model is not reliable.

\smallskip

\begin{tabular*}{\textwidth}{@{\extracolsep{\fill}}lccc}
\\ 
\hline \multicolumn{4}{c}{Panel A: NLL} \\
\hline
 & Violations  &    Percentage       &    LR                   \\
$t$-like$_{95\%}$      &   214  &    4.99\%   &0.00    \\ 
$t$-like$_{99\%}$      &   71  &    1.66\%   &15.55$^{\ast}$    \\ 
Stable-like$_{95\%}$ &   229  &    5.33\%   &    1.02    \\ 
Stable-like$_{99\%}$ &   84  & 1.96\%   &    31.12$^{\ast}$     \\ 
Gaussian$_{95\%}$ &   207  &    4.83\%   &   0.27      \\ 
Gaussian$_{99\%}$ &   69  &    1.61\%   &    13.57$^{\ast}$     \\
\hline \multicolumn{4}{c}{Panel B: NLS} \\
\hline
 & Violations  &    Percentage       &    LR                   \\
$t$-like$_{95\%}$      &   221  &    5.15\%   &0.21    \\ 
$t$-like$_{99\%}$      &   65  &    1.52\%   &9.95$^{\ast}$    \\ 
Stable-like$_{95\%}$ &   242  &    5.64\%   &    3.60   \\ 
Stable-like$_{99\%}$ &   57  &    1.33\%   &    4.26$^{\ast}$  \\ 
Gaussian$_{95\%}$ &   206  &    4.80\%   &   0.35    \\ 
Gaussian$_{99\%}$ &   68  &    1.59\%   &    12.62$^{\ast}$   \\ 
\hline \multicolumn{4}{c}{Panel C: NLDC} \\
\hline
 & Violations  &    Percentage       &    LR                   \\
$t$-like$_{95\%}$      &   244  &    5.69\%   &4.13$^{\ast}$    \\ 
$t$-like$_{99\%}$      &   76  &    1.77\%   &21.01$^{\ast}$    \\ 
Stable-like$_{95\%}$ &   243  &    5.67\%   &    3.86$^{\ast}$    \\ 
Stable-like$_{99\%}$ &   65  & 1.52\%   &    9.95$^{\ast}$     \\ 
Gaussian$_{95\%}$ &   243  &    5.87\%   &   3.86$^{\ast}$    \\ 
Gaussian$_{99\%}$ &   83  &    1.94\%   &    29.77$^{\ast}$     \\ 
\end{tabular*}
}
\end{table}

%%%%%%%meta tables

\begin{table}[t]
\caption{Value-at-Risk backtesting (quadratic approximation)}
\label{tbl:quad-meta2}
{\footnotesize \vspace{2mm} %
\setstretch{1} %\raggedleft
%%
%\scriptsize

This table is a continuation of Table \ref{tbl:quad2}. The same notation is
used here.

\smallskip

\begin{tabular*}{\textwidth}{@{\extracolsep{\fill}}lccc}
\hline \multicolumn{4}{c}{Panel A: NLL} \\
\hline
& Violations  &    Percentage       &    LR                   \\
Meta-$t_{95\%}$ &   224  &    5.22\%   &0.45  \\ 
Meta-$t_{99\%}$ &   51  &    1.19\%   &1.46 \\ 
Meta-stable$_{95\%}$  &   233  &    5.43\%   &1.65   \\ 
Meta-stable$_{99\%}$  &   45  &    1.05\%   &0.10    \\
Meta-$t_{95\%}$ $(\nu_0=\infty)$ &   220  &    5.13\%   &0.15   \\ 
Meta-$t_{99\%}$ $(\nu_0=\infty)$ &   53  &    1.24\%   &2.24    \\ 
Meta-stable$_{95\%}$ $(\alpha_0=2)$ &   221  &    5.15\%   &0.21    \\ 
Meta-stable$_{99\%}$ $(\alpha_0=2)$ &   44  &    1.03\%   &0.03    \\  
\hline \multicolumn{4}{c}{Panel B: NLS} \\
\hline
 & Violations  &    Percentage       &    LR                   \\
Meta-$t_{95\%}$ &   228  &    5.32\%   &0.89  \\ 
Meta-$t_{99\%}$ &   48  &    1.12\%   &0.59  \\ 
Meta-stable$_{95\%}$  &   249  &    5.80\%   &5.60$^{\ast}$    \\ 
Meta-stable$_{99\%}$  &   45  &    1.05\%   &0.10   \\  
Meta-$t_{95\%}$ $(\nu_0=\infty)$ &   227  &    5.29\%   &0.77  \\ 
Meta-$t_{99\%}$ $(\nu_0=\infty)$ &   54  &    1.26\%   &2.69   \\ 
Meta-stable$_{95\%}$ $(\alpha_0=2)$ &   233  &    5.43\%   &1.65    \\ 
Meta-stable$_{99\%}$ $(\alpha_0=2)$ &   45  &    1.05\%   &0.10    \\ 
\hline \multicolumn{4}{c}{Panel C: NLDC} \\
\hline
 & Violations  &    Percentage       &    LR                   \\
Meta-$t_{95\%}$ &   253 &    5.90\%   &6.93$^{\ast}$  \\ 
Meta-$t_{99\%}$ &   66  &    1.54\%   &10.81$^{\ast}$  \\ 
Meta-stable$_{95\%}$  &   262  &    6.11\%   &10.42$^{\ast}$   \\ 
Meta-stable$_{99\%}$  &   56  &    1.31\%   &3.70  \\ 
Meta-$t_{95\%}$ $(\nu_0=\infty)$ &   250  &    5.83\%   &5.92$^{\ast}$   \\ 
Meta-$t_{99\%}$ $(\nu_0=\infty)$ &   66  &    1.54\%   &10.81$^{\ast}$    \\ 
Meta-stable$_{95\%}$ $(\alpha_0=2)$ &   244  &    5.69\%   &4.13$^{\ast}$    \\ 
Meta-stable$_{99\%}$ $(\alpha_0=2)$ &   46  &    1.07\%   &0.22    \\ 
\end{tabular*}
}
\end{table}

%%% endinput

%%% Local Variables: 
%%% mode: latex
%%% TeX-master: "var3"
%%% End: 

\section{Concluding remarks}
Let $G \sim N(0,Q)$, and consider the random vector $X$ obtained from
$G$ by variance-mixture where $S$ is a positive random variable
independent of $G$, that is $X:=S^{1/2}G$. If $S$ has the distribution
of the reciprocal of a $\chi^2$-distributed random variable resp. of
an $\alpha/2$-stable subordinator, we obtain the class of multivariate
$t$ resp. symmetric $\alpha$-stable sub-Gaussian laws. Plenty of other
distributions obtained by variance mixing of a Gaussian measure on
$\erre^d$ that have appeared in the literature for different purposes,
including of course the modeling of financial risk factors. Similarly,
many generalizations of variance mixing have appeared in the
literature, and it is evident that endless variations on the theme are
possible.
In this article we have limited ourselves to two special cases of two
possible extensions. Namely, variance mixture models could be
generalized setting $X=T^{1/2}G$, where $T$ is a positive-definite
random matrix (cf. \cite{BNPA} for related classes of distributions),
or one could consider nonlinear images of (the law of) $S^{1/2}G$,
such as $X=\phi(S^{1/2}G)$, with $\phi$ e.g. a (deterministic)
injective function. In particular, if $T$ is a diagonal random matrix
with independent entries and, for each $i=1,\ldots,d$, $T_{ii}$ is
distributed like the reciprocal of a $\chi^2$ random variable with
$\nu_i$ degrees of freedom, we obtain the class of $t$-like laws of
\S\ref{sec:mtv}. A completely analogous observation holds for the
stable-like laws of \S\ref{sec:stablike}.
Similarly, for particular choices of functions $\phi$, we obtain
meta-elliptical distributions, of which meta-$t$ and meta-stable are
just special cases.

We have considered $t$-like and stable-like laws because of their
simplicity and ease of estimation and simulation, whereas the two
specific instances of meta-elliptical distribution have been
considered for their seemingly widespread use (at least as regards the
meta-$t$ law), especially in connection with applications of copula
methods.

Let us mention other possible generalizations of the classical
variance-mixture approach that have recently appeared in the
literature, without any claim of completeness (which, as already
mentioned above, would not be possible). Assume $X=T^{1/2}G$, with the
same notation as above, where $T$ is diagonal, $T_{ii} =
F_i^\leftarrow(U)$ for each $i=1,\ldots,d$, $U$ is a uniformly
distributed random variable independent of $G$, and and $F_i$,
$i=1,\ldots,d$ are cumulative distribution functions. In the
particular case in which each $F_i$ is the cumulative distribution
function of the reciprocal of a (rescaled) $\chi^2$-distributed random
variable with $\nu_i$ degrees of freedom, we obtain the class of
grouped $t$-distributions (see e.g. \cite{Ban1,Ban2,dema}). Of course
nothing prevents us from considering arbitrary distributions functions
as $F_i$'s.
By combining this construction with a nonlinear map, so that
$X=\phi(T^{1/2}G)$, one could for instance construct laws with the
dependence structure of a grouped-$t$ law and with arbitrary marginals
(see e.g. \cite{dau} for the so-called grouped-$t$ copula). The reader
clearly understand that the possibilities for constructing
multivariate laws by any of these methods, or a combination thereof,
are endless.

\medskip

Our empirical results suggest that, among the infinitely many possible
models for risk factors with non-trivial tail dependence, both classes
of meta-$t$ and meta-stable laws offer good performance, at least on
reasonably diversified portfolios. Nevertheless, there is weaker
evidence that, under ``extreme'' conditions such as those
characterizing our portfolio $B$, these classes of models can perform
well in highly non-linear situations. We believe that the most
important message of our paper is that it is indeed worth taking into
account that ``classical'' multivariate Gaussian laws with changed
marginals (in particular $\alpha$-stable) might perform surprisingly
well. Such (relatively) simple models are undoubtedly attractive from
the viewpoint of practical implementation, as their estimation,
simulation and quadratic approximation are very easy and ``light'' in
terms of computational resources.

\appendix

\section{Densities of some random vectors}
\label{sec:apetta}

\subsection{Densities of a class of images of random vectors}
Let $\phi \in C^1(\erre^d \to \erre^d)$ be an injective function of the type
\[
\phi: (x_1,x_2,\ldots,x_d) \mapsto 
\big( \phi_1(x_1),\phi_2(x_2),\ldots,\phi_d(x_d)\big),
\]
for functions $\phi_k: \erre \to \erre$, $k=1,\ldots,d$.

\begin{prop}
  Let $X$ be a $d$-dimensional random vector with density $p$. Then
  the density of $\phi(X)$ is the function
\[
(y_1,\ldots,y_k) \mapsto \frac{p(\phi_1^{-1}(y_1),\ldots,\phi_d^{-1}(y_d))}%
                 {\phi'_1(\phi_1^{-1}(y_1)) \cdots
                  \phi'_d(\phi_d^{-1}(y_d))}.
\]
\end{prop}
\begin{proof}
  By the multidimensional change of variable formula for Lebesgue
  integrals and by the inverse function theorem we have, for any
  measurable set $A$,
\begin{align*}
  \P(\phi(X) \in A) &= \P(X \in \phi^{-1}(A)) = 
  \int_{\phi^{-1}(A)} p(x)\,dx \\
  &= \int_A p(\phi^{-1}(y))\,\big|\det D\phi^{-1}(y)\big| \,dy \\
  &= \int_A p(\phi^{-1}(y))\,
            \frac{1}{\big|\det D\phi(\phi^{-1}(y))\big|} \,dy \\
  &= \int_A \frac{p(\phi_1^{-1}(y_1),\ldots,\phi_d^{-1}(y_d))}%
                 {\phi'_1(\phi_1^{-1}(y_1)) \cdots
                  \phi'_d(\phi_d^{-1}(y_d))}\,dy_1 \cdots dy_d,
\end{align*}
thus proving the claim.
\end{proof}

\subsection{Densities of sub-Gaussian $\alpha$-stable vectors}
Let $A$ and $G$ be as in Section \ref{sec:metas}, and set
$Y=A^{1/2}G$. Let us recall that the law of $G$ admits the density
\[
\gamma_Q(x) = \frac{1}{(2\pi)^{d/2}} \frac{1}{(\det Q)^{1/2}}
\exp\Big(-\frac12 \ip{Q^{-1}x}{x}\Big).
\]
Therefore, for any constant $c \in \erre$, we have
\[
\gamma_{cQ}(x) = \frac{1}{(2\pi)^{d/2}} \frac{1}{(\det Q)^{1/2}}
c^{-d/2} \exp\Big(-\frac{1}{2c} \ip{Q^{-1}x}{x}\Big).
\]
For any measurable set $B$, recalling that $A$ and $G$ are
independent, we have
\[
\P(Y \in B) = \int_0^\infty \P(a^{1/2}G \in B)p_A(a)\,da
= \int_0^\infty \int_B \gamma_{aQ}(x)\,dx\,p_A(a)\,da,
\]
where $p_A$ denotes the density of the law of $A$. Therefore the law
of $Y$ admits a density $h(x)=g(\ip{Q^{-1}x}{x})$, where
\[
g(z) := \frac{1}{(2\pi)^{d/2}} \frac{1}{(\det Q)^{1/2}}
\int_0^\infty a^{-d/2} e^{-z/(2a)}p_A(a)\,da, \qquad z>0.
\]
In fact $g$ can be extended by continuity at $z=0$, since the above
integral with the exponential term suppressed is well-defined,
e.g. using the series expansion at zero for the density of
$S_{\alpha/2}(1,1,0)$ of \cite[p.~99]{Zolot-1d}.

\section{Prices and sensitivities of some exotic options}
\label{sec:ape}
Throughout this appendix we place ourselves in a standard
Black-Scholes model with one ``underlying'' stock, whose price process
is denoted by $S_t$, $0 \leq t \leq T$, and whose (constant)
volatility is denoted by $\sigma$. The risk-free rate will be denoted
by $r$. We shall consider options written on the stock, denoting the
exercise time by $T$, the strike price by $K$, and the barrier by $H$.

In the following table we collect the definitions, in terms of their
payoff, of some barrier and binary options.
\begin{center}
\begin{tabular}{lll}
\textsc{Name} & \textsc{Payoff} &\\
\hline
Down-and-In call  &  $\max(S_T-K,0)$  &if $\min_{0\leq t\leq T}S_t\leq H$ \\
Down-and-In put &  $\max(K-S_T,0)$ &if $\min_{0\leq t\leq T}S_t \leq H$ \\
Down-and-Out call & $\max(S_T-K,0)$ &if $\min_{0\leq t\leq T}S_t\geq H$\\
Down-and-Out put &  $\max(K-S_T,0)$ &if $\min_{0\leq t\leq T}S_t \geq H$\\
Up-and-In call &  $\max(S_T-K,0)$ &if $\max_{0\leq t\leq T}S_t\geq H$\\
Up-and-In put &  $\max(K-S_T,0)$ &if $\max_{0\leq t\leq T}S_t \geq H$\\ 
Up-and-Out call & $\max(S_T-K,0)$ &if $\max_{0\leq t\leq T}S_t\leq H$\\ 
Up-and-Out put &  $\max(K-S_T,0)$ &if $\max_{0\leq t\leq T}S_t \leq H$\\
Cash-or-Nothing call & $1$ &if $S_T \geq K$\\
Cash-or-Nothing put & $1$ &if $S_T \leq K$
\end{tabular}
\end{center}
We shall use $C_{di}$ and $P_{di}$ to denote the price (at time zero)
of a down-and-in call and a down-and-in put, respectively. Completely
analogous notation will be used for the remaining options, replacing
the subscripts accordingly. The price of plain European call and put
options will be denoted by $C_{BS}$ and $P_{BS}$, respectively. The
price at time zero of a European call option with strike $K$ and
exercise time $T$, written on an underlying whose price at time zero is
$S_0$, will be denoted by $C_{BS}(S_0,K,T)$. The corresponding
notation will be also used for European put options.

Setting
\[
\lambda=\frac{2r}{\sigma^2}-1,
\qquad
m=\frac{r}{\sigma^2}+\frac{1}{2}
\]
and assuming $H<K$, one has (see e.g. \cite{Carr_etal:98}),
\begin{align*}
C_{di} &= H^\lambda S_0^{-\lambda}C_{BS}(H^2S_0^{-1},K,T),\\
P_{di} &= C_{di} + KH^{-1} P_{BS}(S_0,H,T)
          -(HS_0^{-1})^{2m-2} HK^{-1} C_{BS}(KHS_0^{-1},K^2H^{-1},T),\\
C_{ui} &= C_{BS}\\
P_{ui} &= H^\lambda S_0^{-\lambda} P_{BS}(H^2S_0^{-1},K,T).
\end{align*}
By the obvious identities
\[
C_{di} + C_{do} = C_{BS},
\qquad
C_{ui} + C_{uo} = C_{BS},  
\]
and the corresponding ones for put options (i.e. those obtained
replacing $C$ with $P$), we obtain pricing formulas for all barrier
options listed in the above table.
By the well-known formulas for sensitivities of European call and put
options, elementary calculus yields
\begin{align*}
\frac{\partial C_{di}}{\partial S_0} &=
-\lambda H^{\lambda} S_0^{-\lambda-1}C_{BS}(H^2S_0^{-1},K,T)\\
&\quad -H^{\lambda+2} S_0^{-\lambda-2}\Delta_{BS}(H^2S_0^{-1},K,T),\\
\frac{\partial^2 C_{di}}{\partial S_0^2} &=
\lambda(\lambda+1)H^{\lambda}S_0^{-\lambda-2}C_{BS}(H^2S_0^{-1},K,T)\\
&\quad +2(\lambda+1)H^{\lambda+2}S_0^{-\lambda-3}\Delta_{BS}(H^2S_0^{-1},K,T)\\
&\quad +H^{\lambda+4}S_0^{-\lambda-4}\Gamma_{BS}(H^2S_0^{-1},K,T),\\
\frac{\partial C_{di}}{\partial T} &=
H^\lambda S_0^{-\lambda}\Theta_{BS}(H^2S_0^{-1},K,T).
\end{align*}
Similar expressions can be derived for the sensitivities of the other
binary options.

Setting 
\[
d_1 := \frac{\log(S_0/K)+(r+\sigma^2)T}{\sigma\sqrt{T}},
\qquad
d_2 := d_1-\sigma\sqrt{T},
\]
we have (see e.g. \cite{hull})
\[
C_{cn} = e^{-rT} \Phi(d_2),
\qquad
P_{cn} = e^{-rT} \Phi(-d_2),
\]
where $\Phi(\cdot)$ stands for the distribution function of the
Gaussian law on $\erre$ with mean zero and unit variance.  The
sensitivities of binary options are just an exercise in elementary
calculus. Let us include, for the sake of completeness, the
sensitivities of the cash-or-nothing put, which is used in our
portfolios:
\begin{align*}
\frac{\partial P_{cn}}{\partial S_0} &=
\frac{-e^{-rT}\Phi'(-d_2)}{\sigma S_0\sqrt{T}},\\
\frac{\partial^2 P_{cn}}{\partial S_0^2} &=
\frac{e^{-rT}\Phi'(-d_2)}{\sigma S_0^2\sqrt{T}}
+\frac{-d_2 e^{-rT-d_2^2/2}}{\sigma^2 T S_0^2 \sqrt{2\pi}},\\
\frac{\partial P_{cn}}{\partial T} &=
- r e^{-rT} \Phi(-d_2)
+ \frac{r e^{-rT} \Phi'(-d_2) \log(S_0/K)}{2\sigma T^{3/2}} 
- \frac{r-\sigma^{2}/2}{\sigma\sqrt{T}}.
\end{align*}

\bibliographystyle{amsplain}
\bibliography{ref,var1}

\end{document}